\documentclass[10pt]{article}
\usepackage{amsmath,amssymb,cite,epsfig}
\usepackage{graphics}
\def \beq{\begin{equation}}
\def \eeq{\end{equation}}
\def \beqa{\begin{eqnarray}}
\def \eeqa{\end{eqnarray}}
\def \la{\langle}
\def \ra{\rangle}
\def \l{\left(}
\def \r{\right)}
\def \Lag{\mathcal{L}}
\newcommand{\nn}{\nonumber}
\newcommand{\Tr}{{\mathrm{Tr}}}
\newcommand{\fk}{\ensuremath{f_{K}}}
\newcommand{\bsig}{\ensuremath{\bar{\sigma}}}
\usepackage{xspace}

\begin{document}

% \title{Including the Fermion Vacuum Fluctuations in the $(2+1)$ flavor Polyakov Quark Meson Model}
% \author{Sandeep\ \surname{Chatterjee}}
% \email{sandeep@cts.iisc.ernet.in}
% \affiliation{Centre for High Energy Physics,\\ Indian Institute of Science,\\
%         Bangalore 560012, India.}
% \author{Kirtimaan A.\ \surname{Mohan}}
% \email{kirtimaan@cts.iisc.ernet.in}
% \affiliation{Centre for High Energy Physics,\\ Indian Institute of Science,\\
%          Bangalore 560012, India.}

\title{\begin{flushright}
% \small IISc/CHEP/11/x
\end{flushright}
%\vspace{0.25cm} 
{\bf Including the Fermion Vacuum Fluctuations in the $(2+1)$ flavor Polyakov Quark Meson Model}}
\author{Sandeep Chatterjee \footnote{sandeep@cts.iisc.ernet.in} \,and Kirtimaan A. Mohan\footnote{kirtimaan@cts.iisc.ernet.in}\\ 
\begin{small}
Centre for High Energy Physics, Indian Institute of Science, Bangalore, 560012, India.
\end{small}}
\date{\empty}
\maketitle

\begin{abstract}

We consider the $(2+1)$ flavor Polyakov Quark Meson Model and study the effect of 
including fermion vacuum fluctuations on the thermodynamics and phase diagram. The resulting 
model predictions are compared to the recent QCD lattice simulations by the HotQCD and 
Wuppertal-Budapest collaborations. The variation of the thermodynamic quantities across 
the phase transition region becomes smoother. This results in better agreement with the lattice data.
Depending on the value of the mass of the sigma meson, including the vacuum term 
results in either pushing the critical end point into higher values of the chemical potential or 
excluding the possibility of a critical end point altogether. 
\end{abstract}

% \pacs{12.38Aw, 12.39.-x, 12.38.Gc, 05.70.-a}

\maketitle

\section{Introduction}

With the advent of ultrarelativistic heavy ion experiments such as RHIC (BNL), LHC (CERN)
and the planned future CBM experiment at the FAIR facility in Darmstadt, research on the properties
of strongly interacting matter has entered into a fascinating era. Quantum Chromodynamics (QCD)
which is the theory of strong interactions predicts that at high temperature and baryon density normal
hadronic matter undergoes phase transition to a chirally symmetric phase of Quark Gluon Plasma (QGP)~\cite{QGP1,QGP2,QGP3,QGP4,QGP5}.

QCD at this temperature and density is a strongly coupled theory and hence one cannot use perturbative
methods to study the details of this phase transition. Non-perturbative methods like first principle
Lattice QCD (LQCD) Monte Carlo simulations give us important insights into various aspects of the
phase transition. However LQCD suffers from the notorious sign problem at non zero baryon
density~\cite{LQCD-sign1,LQCD-sign2}. Although several methods have been developed~\cite{LQCD-sign1,LQCD-sign2}
to bypass the sign problem at small baryon chemical potential $\mu_B$, a satisfactory solution
to the sign problem for all values of $\mu_B$ still eludes us. An alternate approach is to study various 
phenomenological models whose phase diagrams possess the essential features of QCD. These models 
serve to complement LQCD calculations and may provide us with an intuitive understanding about
regions of the phase diagram currently inaccessible to LQCD.

The QCD Lagrangian for $N_f$ flavors of massless quarks has the global symmetry 
$SU_{L}\l N_f\r\times SU_{R}\l N_f\r$ which spontaneously breaks into $SU_{V}\l N_f\r$ in the low 
energy hadronic vacuum by the formation of chiral condensate together with $N_f^2-1$ massless Goldstone bosons. 
In the opposite limit of infinite quark mass, QCD becomes a pure $SU\l3\r$ gauge theory. The low energy 
vacuum that possesses a center symmetry  $Z\l3\r$ under the color gauge group (confined phase) 
gets spontaneously broken in the high temperature/baryon density regime (deconfined phase).
The Polyakov loop which is vanishing in the confining phase and becomes non zero in the symmetry broken 
phase serves as the order parameter of the confinement-deconfinement phase transition~\cite{Polyakov}. In real 
life we have dynamical quarks with non zero masses and hence neither of the symmetries discussed above are exact. 
Thus the chiral condensate and the Polyakov loop only serve as approximate order parameters for the chiral 
phase transition and confinement-deconfinement phase transition respectively.

There have been numerous attempts to study the QCD phase diagram using effective chiral models like the 
Nambu-Jona-Lasinio (NJL)~\cite{NJL1,NJL2,NJL3,NJL4}, linear sigma~\cite{LSM1,LSM2,lenaghan} and the quark-meson 
(QM)~\cite{QM} models. These models possess QCD like chiral symmetry breaking pattern. Later, variants of these were 
developed to incorporate the confinement-deconfinement transition, namely PNJL~\cite{PNJL1,PNJL2,PNJL3,PNJL4}, 
PLSM and PQM~\cite{Schaefer:07,H.mao09} models. There is a basic difference in the mechanism of chiral symmetry breaking 
in the QM/PQM models compared to NJL/PNJL models. In the latter case, the fermion vacuum term which is the contribution 
due to the infinite Dirac fermion sea causes the symmetry breaking. In QM/PQM models, the spontaneous symmetry breaking 
is generated by the mesonic potential and the fermion vacuum term is usually neglected. Recently in the 2 flavor PQM model
it has been shown that the order of the phase transition in the massless limit changes from first order to crossover on 
adding the vacuum term~\cite{PQMVT-Skokov} which also modifies the shape of the isentropic trajectories near 
the critical end point (CEP)~\cite{PQMVT-isen}. In yet another study in the presence of a magnetic field~\cite{PQMVT-mag}, 
the phase diagram is shown to be considerably affected by the vacuum term. More recently, the effect of the vacuum term in 
the 2 flavor case on the thermodynamics as well as the phase diagram was extensively studied~\cite{PQMVT-USG}. 

In this paper we will consider the PQM model together with the vacuum term (PQMVT) in the $(2+1)$ flavor case. We first
compute the thermodynamic potential in the presence of the vacuum term. Then we calculate thermodynamic quantitites
like energy density $\epsilon$, pressure $p$, entropy $s$, conformal symmetry breaking measure $\Delta$, speed of
sound $c_s^2$ and specific heat capacity $c_V$. We also compute several second order quark number susceptibilities (QNS).
In each case we compare our results with those obtained from LQCD and comment on the modification in the model prediction due 
to the addition of the vacuum term. We also present the phase diagram of PQMVT in the $T-\mu_B$ plane. The organisation of the
paper is as follows: In Sec.~\ref{sec.model}, we discuss the details of the PQMVT model and its parameters. In Sec.~\ref{sec.results}
we present our results and elucidate the effect of the vacuum term on thermodynamics and phase diagram of PQMVT. Finally, we
summarise and conclude in Sec.~\ref{sec.conc}.

\section{The Model}
\label{sec.model}

We will be working in the generalized three flavor quark meson linear 
sigma model which has been combined with the Polyakov loop potential~
\cite{Schaefer:09wspax,H.mao09,PQMlat}. 
In this model, quarks denoted by $q_f=\l u,d,s\r^T$ come in $N_f=3$ flavor and 
$N_c=3$ color degrees of freedom and are coupled to the $SU_L(3) \times SU_R(3)$ 
symmetric mesonic fields $M$ as well as a spatially constant temporal gauge field 
$A^{\mu}$.
Our starting point is the following model Lagrangian
\beq
 \Lag=\Tr \left( \partial_\mu M^\dagger \partial^\mu M \right)+
  \bar{q_f}\big( i \gamma^{\mu}D_{\mu}-gM_5 \big) q_f-{\cal U}_M\l M\r
  -{\cal U}_P\l\Phi,\bar{\Phi},T\r,
 \label{lagrangian}
\eeq
where nine mesons in the scalar ($\sigma_a, J^{P}=0^{+}$) and 
pseudoscalar ($\pi_a, J^{P}=0^{-}$) sectors are together put into 
the $3 \times 3$ complex matrices $M$ and $M_5$ given by
\beqa
\label{eq:Mfld}
M &=& T_a \xi_a = T_a(\sigma_a +i\pi_a)\\
M_5 &=& T_a \xi_{5a} = T_a(\sigma_a +i\gamma_5\pi_a)
\eeqa
where $T_a=\frac{\lambda_a}{2}$ represent the 9 generators of $U(3)$ with 
$\lambda_0=\sqrt{\frac{2}{3}}\ \bf 1$. The generators follow $U(3)$ algebra 
$\left[T_a, T_b\right]  = if_{abc}T_c$ 
and $\left\lbrace T_a, T_b\right\rbrace  = d_{abc}T_c$ where 
$f_{abc}$ and $d_{abc}$ are standard antisymmetric and symmetric 
structure constants respectively with 
$f_{ab0}=0$ and $d_{ab0}=\sqrt{\frac{2}{3}}\ \bf 1 \ \delta_{ab}$ 
and are normalized as $\Tr(T_a T_b)=\frac{\delta_{ab}}{2}$.

In (\ref{lagrangian}), ${\cal U}_M\l M\r$ and ${\cal U}_P\l\Phi,\bar{\Phi},T\r$ are the
mesonic and Polyakov loop potentials respectively.
Polyakov loop field $\Phi(\vec{x})$ is defined as the thermal 
expectation value of color trace of the Wilson loop in the temporal direction 
\beq
\Phi(\vec{x}) = \frac{1}{N_c}\langle\Tr_c L\l\vec{x}\r\rangle,
\eeq
where $L\l\vec{x}\r$ is a matrix of the 
$SU_c(3)$ color gauge group.
\beq
\label{eq:Ploop}
L(\vec{x})=\mathcal{P}\mathrm{exp}\left[i\int_0^{\beta}d \tau
A_0(\vec{x},\tau)\right]
\eeq
Here $\mathcal{P}$ denotes path ordering, $A_0$ is the temporal vector field and 
$\beta = T^{-1}$~\cite{Polyakov}.
Also, $\bar{\Phi}=\frac{1}{N_c}\langle\Tr_c L^{\dagger}\l\vec{x}\r\rangle$
is the Hermitean conjugate of $\Phi$. The coupling of quarks with the gauge 
field is implemented through the covariant derivative 
$D_{\mu} = \partial_{\mu} -i A_{\mu}$ 
and  $A_{\mu} = \delta_{\mu 0} A_0$ (Polyakov gauge), where 
$A_{\mu} = g_s A^{a}_{\mu} \lambda^{a}/2$. $\lambda_a$ are the Gell-Mann matrices 
in the color space where the index $a$ runs from $1 \cdots 8$ and $g_s$ is the $SU_c\l3\r$ 
gauge coupling. g is the flavor blind Yukawa coupling that couples the three 
quark flavors with the mesons $\sigma_a$ and $\pi_a$ of the meson matrix $M_5$.

The mesonic potential $\mathcal{U}_M\l M\r$ has the following form
\beqa  
\label{eq:Lagmes}
 \mathcal{U}_{M}\l M\r & = & m^2 \Tr ( M^\dagger M) +
  \lambda_1 \left[\Tr (M^\dagger M)\right]^2 \nn \\
  && + \lambda_2 \Tr\left(M^\dagger M\right)^2
  -c   \big[det (M) + det (M^\dagger) \big] \nn \\
  && - \Tr\left[H(M + M^\dagger)\right].
\eeqa

While the 't Hooft determinant term breaks explicitly the $U\l1\r_A$
symmetry, the $SU_L(3) \times SU_R(3)$ chiral symmetry is explicitly 
broken by $H=T_ah_a$ and the $\xi$ field picks up a non zero vacuum 
expectation value $\bar{\xi}$. In order to give the correct vacuum quantum 
numbers after chiral symmtery breakdown, $h_0$, $h_3$ and $h_8$ are the
only allowed non zero parameters in $H$. In this work we retain isospin symmetry which
further sets $h_3=0$. This leads to the $2+1$ flavor symmetry breaking scenario with 
nonzero condensates $\bar{\sigma_0}$ and $\bar{\sigma_8}$.

The effective Polyakov loop potential ${\cal U}_P \left( \Phi, \bar{\Phi},
T \right)$ is constructed to reproduce the thermodynamics of pure gauge
theory on the lattice. The simplest $Z(3)$ symmetric polynomial form based on a Ginzburg-Landau
ansatz is given by~\cite{PNJL3}
\beq
 \frac{{\cal U_{\text{Poly}}}\left(\Phi,\bar{\Phi}, T \right)}{T^4} =
  -\frac{b_2(T)}{2}\Phi\bar{\Phi}-
  \frac{b_3}{6}\l\Phi^3+{\bar{\Phi}}^3\r+\frac{b_4}{4}\l\Phi\bar{\Phi}\r^2.
\label{eq.polpot}
\eeq
Later an improved ansatz was proposed~\cite{rattiimproved,VM} by including the VanderMonde term which
is the Jacobian of transformation from the matrix valued field $L$ to the complex valued field $\Phi$.
Here we closely follow~\cite{VM}
\beq
 \frac{{\cal U_{\text{Poly-VM}}}\left(\Phi,\bar{\Phi}, T \right)}{T^4} =
 \frac{{\cal U_{\text{Poly}}}\left(\Phi,\bar{\Phi}, T \right)}{T^4} -
  \kappa \log\left[1-6\Phi\bar{\Phi}+4\l\Phi^3+{\bar{\Phi}}^3\r-3\l\Phi\bar{\Phi}\r^2\right].
\label{eq.VMpot}
\eeq
The logarithm term constrains the $\Phi$ and $\bar{\Phi}$ values to smaller than 1.
Details of the above mentioned model parameters will be discussed in Section~\ref{parameters}.
\subsection{Grand Potential}

The grand canonical partition function for a spatially uniform system in 
thermal equilibrium at finite temperature T and quark chemical potential
$\mu_f (f=u, d, s)$ is given by
\beqa
\mathcal{Z}&=& \mathrm{Tr\, exp}[-\beta (\hat{\mathcal{H}}-\sum_{f=u,d,s} 
\mu_f \hat{\mathcal{N}}_f)] \nn \\
&=& \int\prod_a \mathcal{D} \sigma_a \mathcal{D} \pi_a \int
\mathcal{D}q \mathcal{D} \bar{q} \; \mathrm{exp} \bigg[- \int_0^{\beta}d\tau\int_Vd^3x
\bigg(\mathcal{L^{E}} 
 + \sum_{f=u,d,s} \mu_{f} \bar{q}_{f} \gamma^0 q_{f} \bigg) \bigg]. 
\label{eq.Z}
\eeqa
where V is the three dimensional volume of the system, and 
$\beta= \frac{1}{T}$. In general, the quark chemical potentials 
corresponding to the three conserved flavors up(u), down(d) 
and strange(s) are different. In the $2+1$ flavor setup,
the $SU_V(2)$ symmetry is unbroken and thus the quark chemical 
potentials for u and d quarks become equal $\mu_u = \mu_d=\mu_x $
while the strange quark chemical potential is $\mu_s = \mu_y$.

We evaluate $\mathcal{Z}$ in the mean-field approximation 
\cite{Mocsy:01prc,Schaefer:09}. In this approximation,
thermal and quantum fluctuations of the meson fields are neglected and 
are replaced by their expectation values
$\langle M \rangle =  T_0 \la\sigma_0\ra + T_8 \la\sigma_8\ra$ 
while quarks and antiquarks are retained as quantum fields.
Now following the standard procedure~\cite{Kapusta_Gale,Schaefer:07,PNJL2,PNJL3}
one can obtain the expression of the grand potential as a sum of a pure 
gauge field contribution ${\cal U_{\text{P}}} \l\Phi, \bar{\Phi}, T \r$, 
a meson contribution ${\cal U_{\text{M}}} \left(\langle M\rangle\right)$ 
and a quark-antiquark contribution $\Omega_{\bar{q}q}$
\beq
 \Omega \l T, \mu_x,\mu_y\r = - \frac{T\ln Z}{V} = {\cal U_{\text{M}}}\l\la M\ra\r +
 {\cal U_{\text{P}}} \l\Phi, \bar{\Phi}, T \r+ \Omega_{\bar{q}q}
\label{eq.omega}
\eeq

In order to study the 2 + 1 flavor case, a more convenient basis is
the nonstrange-strange (x, y) basis obtained from the original
singlet-octet (0,8) basis by the following basis transformation,
\beq
\l\begin{array}{c}v_x\\v_y\end{array}\r=\frac{1}{\sqrt{3}}\l
\begin{array}{cc}\sqrt{2} & 1\\1 & -\sqrt{2}\end{array}\r
\l\begin{array}{c}v_0\\v_8\end{array}\r
\eeq
applied on the scalar, pseudoscalar mesons and external fields, that
is $v\in \{\sigma,\pi,h\}$. In this new basis the nonstrange 
and strange quarks and antiquarks decouple and the quark masses become
\beq
m_x = g \frac{\sigma_x}{2}, \qquad m_y = g \frac{\sigma_y}{\sqrt{2}}
\label{quarkmass}
\eeq

The mesonic potential in this nonstrange-strange basis reads,
\beqa
\label{eq:mesop}
 U(\sigma_{x},\sigma_{y})&=&{\cal U_{\text{M}}}\l\la M\ra\r\nn\\
  &= &\frac{m^{2}}{2}\left(\sigma_{x}^{2} +
  \sigma_{y}^{2}\right) -h_{x} \sigma_{x} -h_{y} \sigma_{y}
 - \frac{c}{2 \sqrt{2}} \sigma_{x}^2 \sigma_{y} \nn \\
 && + \frac{\lambda_{1}}{2} \sigma_{x}^{2} \sigma_{y}^{2}+
  \frac{1}{8}\left(2 \lambda_{1} +
    \lambda_{2}\right)\sigma_{x}^{4} \nn \\
 && +\frac{1}{8}\left(2 \lambda_{1} +
    2\lambda_{2}\right) \sigma_{y}^{4}\ ,
\eeqa    
while the quark-antiquark contribution is given by,
\beq
\Omega_{\bar{q}q} =\Omega^{\text{v}}_{\bar{q}q}+
 \Omega^{\text{th}}_{\bar{q}q}
\label{omegaqq}
\eeq
where
\beqa
\Omega^{\text{v}}_{\bar{q}q}&=&-2N_c\sum_{f=u,d,s}\int \frac{d^3 p}{(2\pi)^3}E_f \label{ov}\\
\Omega^{\text{th}}_{\bar{q}q}&=&-2T \sum_{f=u,d,s} \int \frac{d^3 p}{(2\pi)^3}
\Bigl[ \ln g_{f}^{+} + \ln g_{f}^{-} \Bigr]
\label{omegafer}
\eeqa
$g_{f}^{+}$ and $g_{f}^{-}$ are defined after taking trace over color space
\beqa
 g_{f}^{+}&=& \Big[ 1 + 3\Phi e^{ -E_{f}^{+} /T} +3 \bar{\Phi}e^{-2 E_{f}^{+}/T} +e^{-3 E_{f}^{+} /T}\Big]\nn\\
 g_{f}^{-}&=& \Big[ 1 + 3\bar{\Phi} e^{ -E_{f}^{-} /T} +3 \Phi e^{-2 E_{f}^{-}/T} +e^{-3 E_{f}^{-} /T}\Big]
\label{eq.gf}
\eeqa
Here we use the notation E$_{f}^{\pm} =E_f \mp \mu_f $ where $E_f$ is the
single particle energy of a quark/antiquark.
\beq
E_f = \sqrt{p^2 + m{_f}{^2}}
\eeq
At zero temperature and chemical potentials, $\Omega_{\bar{q}q}$
gets a contribution only from $\Omega^{\text{v}}_{\bar{q}q}$. This is the 
fermion vacuum contribution which is usually neglected. We shall study the 
effect of this term on the thermodynamics. Employing dimensional regularisation
to regularise the diverging integral in (\ref{ov}), we obtain~\cite{PQMVT-Skokov,PQMVT-USG}
\beq
\Omega^{\text{v}}_{\bar{q}q}=\Omega_{\bar{q}q}^{\text{reg}}\l\Lambda\r=
 -\frac{N_c}{8 \pi^2}\sum_{f=u,d,s}m_f^4\log\left[\frac{m_f}{\Lambda}\right]
\label{eq.omegareg}
\eeq
where $\Lambda$ is the regularisation scale parameter.
In Appendix \ref{modelparameters}, we show that the thermodynamic potential is independent of $\Lambda$. 
Thus all physical observables are independent of the choice of $\Lambda$~\cite{PQMVT-Skokov}.

The quark condensates $\la\sigma_x\ra$, $\la\sigma_y\ra$ and the Polyakov
loop expectation values $\la\Phi\ra$, $\la\bar{\Phi}\ra$ are obtained by 
searching for a global minima of the grand potential at a given value of 
temperature T and chemical potentials $\mu_x$ and $\mu_y$,

\beq
  \left.\frac{ \partial \Omega}{\partial
      \sigma_x} = \frac{ \partial \Omega}{\partial \sigma_y} 
      = \frac{ \partial \Omega}{\partial \Phi}
      = \frac{\partial \Omega}{\partial \bar{\Phi}}
  \right|_{\sigma_x = \langle\sigma_x\rangle, \sigma_y=\langle\sigma_y\rangle,
   \Phi=\langle\Phi\rangle, \bar{\Phi} =\langle\bar{\Phi}\rangle} = 0\ .
\label{eq.gapeq}
\eeq
\subsection{Model Parameters}
\label{parameters}
The model parameters are fixed by inputs from lattice and experiments. 
The parameters of the mesonic potential namely 
$m^2$, $\lambda_1$, $\lambda_2$, $c$, $g$, $h_x$ and $h_y$ are fixed by using the 
following experimentally known quantities as in \cite{lenaghan,Schaefer:09}: 
pion mass $m_{\pi}$ and pion decay constant $f_\pi$, kaon mass $m_K$ and kaon 
decay constant $f_K$, average squared mass of $\eta$ and $\eta'$ mesons, 
$\l m_{\eta}^2+m_{\eta'}^2\r$ and sigma mass $m_\sigma$. $g$ is fixed by using a 
light quark constituent mass $m_x=300$ MeV in (\ref{quarkmass}).
This gives a strange quark constituent mass $m_s \approx 433$ MeV. It turns out
that $\lambda_2$ is the only parameter that depends on $\Lambda$. However, this
dependence cancels with that of $\Omega^{\text{reg}}_{\bar{q}q}\l\Lambda\r$
to yield a $\Lambda$ independent thermodynamic potential. Thus, $\Lambda$ is an arbitrary
parameter here and we do not need any further experimental input to fix it.
In Appendix \ref{modelparameters} we discuss in detail the fixing of these parameters. 
Experimentally the mass of the $\sigma$ meson $m_{\sigma}$ lies in the range
$400-1200$ MeV~\cite{PDG} while recent studies point to $m_{\sigma}\sim 400-500$ MeV~\cite{msig1}. 
All our results are obtained with $m_{\sigma} = 400$ MeV.
In Table \ref{tb.param}, we have given the parameters used in this work.

We now discuss the parameters of the gauge potentials  $\mathcal{U}_{\text{Poly}}(\Phi,\bar{\Phi},T)$ and 
$\mathcal{U}_{\text{Poly-VM}}(\Phi,\bar{\Phi},T)$ used in this work.
\beqa
\frac{{\cal U_{\text{Poly-VM}}}\left(\Phi,\bar{\Phi}, T \right)}{T^4} &=&
 \frac{{\cal U_{\text{Poly}}}\left(\Phi,\bar{\Phi}, T \right)}{T^4} -
  \kappa \log\left[1-6\Phi\bar{\Phi}+4\l\Phi^3+{\bar{\Phi}}^3\r-3\l\Phi\bar{\Phi}\r^2\right]
\label{eq.logpot2}
\eeqa
where $\cal U_{\text{Poly}}$ is given by
\beq
 \frac{{\cal U_{\text{Poly}}}\left(\Phi,\bar{\Phi}, T \right)}{T^4} =
  -\frac{b_2(T)}{2}\Phi\bar{\Phi}-
  \frac{b_3}{6}\l\Phi^3+{\bar{\Phi}}^3\r+\frac{b_4}{4}\l\Phi\bar{\Phi}\r^2.
\label{eq.polpot2}
\eeq
Here $b_2\l T\r$ is given by
\beq
b_2\l T\r=a_0+a_1\frac{T_0}{T}+a_2\l\frac{T_0}{T}\r^2+a_3\l\frac{T_0}{T}\r^3
\label{b2}
\eeq
The parameters of $b_2(T)$, $b_3$ and $b_4$ are obtained by fitting to the
pure gauge lattice data in~\cite{PNJL3}.
\begin{eqnarray}
&& a_0 = 6.75\ , \qquad a_1= -1.95\ , \qquad b_3= 0.75 \nn \\ 
&& a_2 =2.625\ ,  \qquad a_3=-7.44\ ,   \qquad b_4=7.5 \nn 
\end{eqnarray}
The remaining parameters $T_0$ and $\kappa$ are fixed by comparing the model predictions to lattice data 
of the full $(2+1)$ QCD. $T_0=270$ MeV is the phase 
transition temperature in the pure gauge theory. In the presence of dynamical quarks, fermionic contributions modify 
the running coupling of QCD~\cite{running1,running2} which can be accounted for by a smaller value of $T_0$~\cite{running1,running2,Schaefer:07}. 
While both HotQCD as well as WB groups agree that the chiral 
symmetry restoration takes place through a smooth 
crossover~\cite{LQCD-crossover1,LQCD-crossover2,LQCD-crossover3,LQCD-crossover4}, 
there is still a disagreement on the relative values 
of $T_{\chi}$ and $T_d$, the transition temperatures associated with chiral and confinement-deconfinement crossovers
respectively. The HotQCD lattice data~\cite{Lat-Cheng,Lat-Baz} with which we compare our model predictions 
show a coincidence of both transitions. On the other hand, WB sees $T_{\chi}<T_d$~\cite{LQCD-crossover3, Lat-WB, LQCD-WB1, LQCD-WB2}. 
However, recently HotQCD have 
reported lower value of $T_{\chi}$~\cite{Hotlatnew} which is in agreement with that of WB.  
We observe that by varying $T_0$ it is possible to change the relative values of $T_{\chi}$ and $T_{d}$ in the model.
We work with two values of $T_0$: firstly, $T_0=270$ MeV. In this case $T_{\chi}<T_d$. We call this ModelWB. 
Secondly, $T_0=210$ MeV. In this case $T_{\chi}\simeq T_d$. We call this ModelHotQCD. The remaining parameter
$\kappa$ is adjusted to describe the lattice data of the scaled conformal symmetry breaking measure $\Delta/T^4 = \l \epsilon-3p\r/T^4$. 
The two lattice groups differ in the height of the peak of $\Delta$~\cite{Lat-WB}: the peak height is about 
50 $\%$ larger in HotQCD. Here we use different values of $\kappa$ in ModelWB and ModelHotQCD to describe the WB and
HotQCD data respectively. One could have obtained
$T_0$ and $\kappa$ by precision fitting of the LQCD data. Here we do not undertake such a procedure as the
exact quantitative status is going to change as more refined LQCD data become available. Here we merely focus
on comparing the trends between model and LQCD rather than exact matching. We summarise below the values of all the model 
parameters in Table \ref{tb.param}.

%%%%%%%%%%%%%%%%%%%%%%%%%%%%%%%%%%%%%%%%%%%%%%%%%%%%%%%%%%%%%%%%%%%%%%%%%%%%%%%%
%             Table: Parameter Set

\begin{table}[htb]
\begin{center}
\begin{tabular}{|c|c|c|c|c|c|c|c|c|c|c|}
\hline
Model&$m_{\sigma}$ $[\text{MeV}]$&$m^2$ $[\text{MeV}^2]$&$\lambda_1$&$\lambda_2\l\Lambda\r$&$c$&$h_x$ 
$[\text{MeV}^3]$&$h_y$ $[\text{MeV}^3]$&$g$&$T_0$ [MeV]&$\kappa$\\
\hline
\hline
ModelHotQCD&400&80647.587&-8.165&138.45&4801.95&$1.785\times10^6$&$3.805\times10^7$&6.5&210 &0.1 \\
\hline
ModelWB&400&80647.587&-8.165&138.45&4801.95&$1.785\times10^6$&$3.805\times10^7$&6.5&270 &0.2 \\
\hline
\end{tabular}
\end{center}
\caption{The parameter sets obtained with $\Lambda=200$ MeV.}
\label{tb.param}
\end{table}

% %%%%%%%%%%%%%%%%%%%%%%%%%%%%%%%%%%%%%%%%%%%%%%%%%%%%%%%%%%%%%%%%%%%%%%%%%%%%%%%%
% %             Table: Parameter Set
% 
% \begin{table}[htb]
% \begin{center}
% \begin{tabular}{|c|c|c|c|c|c|c|c|}
% \hline
% Model&$T_0$ MeV&$\,c\,$&$\,d\,$&$T_{\chi}$ MeV&$T_d$ MeV\\
% \hline
% \hline
% ModelHotQCD&210&0.1&2&179&174\\
% \hline
% ModelWB&270&0.2&2&207&240\\
% \hline
% \end{tabular}
% \end{center}
% \caption{Values of the parameters $T_0$, $c$ and $d$ and the transition temperatures $T_{\chi}$ and $T_d$ 
% obtained in ModelHotQCD and ModelWB.}
% \label{tb.uparam}
% \end{table}
% %%%%%%%%%%%%%%%%%%%%%%%%%%%%%%%%%%%%%%%%%%%%%%%%%%%%%%%%%%%%%%%%%%%%%%%%%%%%%%%%

\section{Results}
\label{sec.results}

\begin{figure}
 \begin{center}
\scalebox{0.75}{\includegraphics{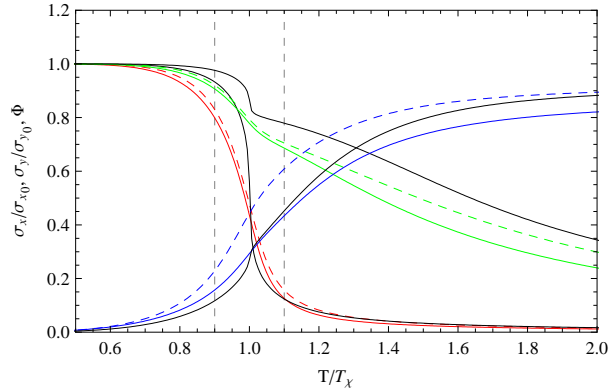}}
\end{center}
\caption{Plots of normalised $\langle\sigma_x\rangle$(red), $\langle\sigma_y\rangle$
 (green) and $\langle\Phi\rangle$(blue) vs $\displaystyle{T/T_{\chi}}$ at $\mu_x=\mu_y=0$ as obtained
in PQMVT with $\mathcal{U}_{\text{Poly-VM}}$ as the Polyakov potential. The dashed curves 
refer to ModelHotQCD while the solid curves refer to ModelWB. The solid black curve in each case
is obtained in PQM with the ModelWB parameter set.}
\label{fg.ordparam}
\end{figure}

Having obtained the grand canonical potential $\Omega\l T,\mu_x,\mu_y\r$ in the presence of the 
vacuum term, we shall now compute various thermodynamic quantities and study the effect of 
inclusion of the vacuum term on these quantities.
For a given $\l T,\mu_x,\mu_y\r$, at the mean field level, $\Omega\l T,\mu_x,\mu_y\r$ is only a function of 
the condensates $\sigma_x$, $\sigma_y$, $\Phi$ and $\bar{\Phi}$ which are determined by solving the gap 
equations (\ref{eq.gapeq}) simultaneously. In Fig.~(\ref{fg.ordparam}) 
we have plotted the condensates $\langle\sigma_x\rangle$, $\langle\sigma_y\rangle$ and $\langle\Phi\rangle$ 
obtained in ModelHotQCD and ModelWB at zero chemical potentials ($\mu_x=\mu_y=0$) with $\mathcal{U}_{\text{Poly-VM}}$ as 
the Polyakov potential. We do not plot $\langle\bar{\Phi}\rangle$ since at zero chemical potential 
$\langle\bar{\Phi}\rangle=\langle\Phi\rangle$. While the quark condensates
come out similar in both the models, using a higher value of $T_0$ and $\kappa$ in ModelWB results in a delayed 
confinement-deconfinement crossover as well as a suppressed value for the Polyakov condensate as compared to
ModelHotQCD. For large $T$, $\langle\Phi\rangle$ in ModelWB falls short by $ 8\%$ compared to that of ModelHotQCD. In
order to compare the results between PQM and PQMVT, we have also plotted the results as obtained in PQM
with the ModelWB parameter set. In PQM, $\langle\sigma_x\rangle$ drops 
sharply across the transition region accompanied by an unsmooth structure in $\langle\sigma_y\rangle$ 
and $\langle\Phi\rangle$. With the inclusion of the vacuum term, the drop in $\langle\sigma_x\rangle$ across 
the crossover is much more gentle and the jagged structures in the transition region in $\langle\sigma_y\rangle$ 
and $\langle\Phi\rangle$ are washed away. One can compare these condensates as obtained in PQMVT with
those computed in LQCD.

\begin{figure}
\begin{center}
\scalebox{0.75}{\includegraphics{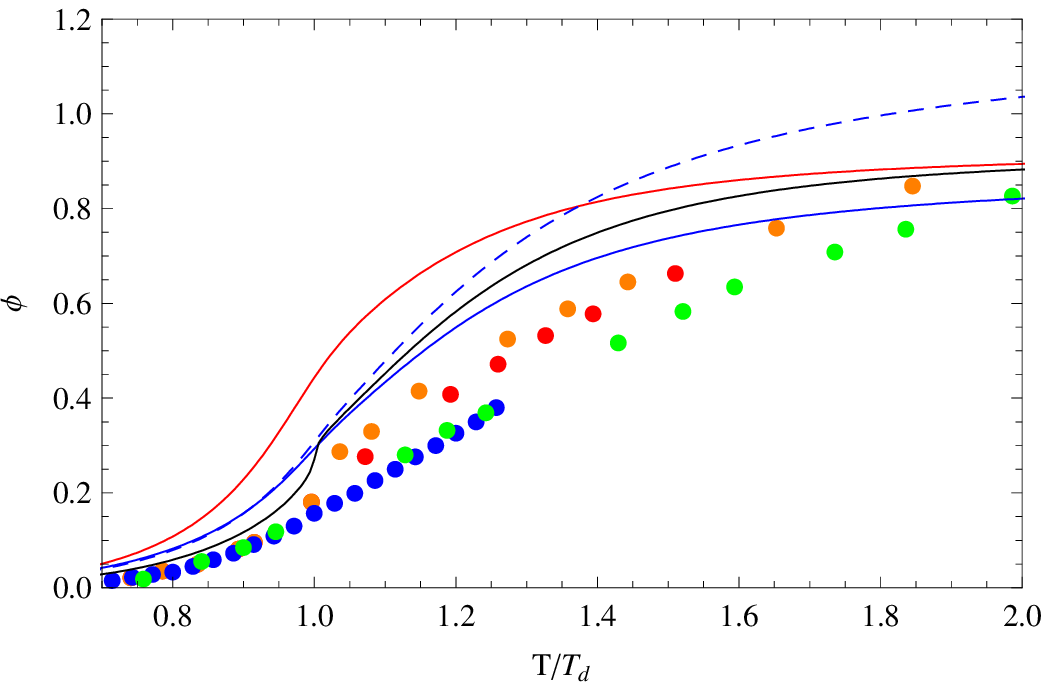}}
\scalebox{0.75}{\includegraphics{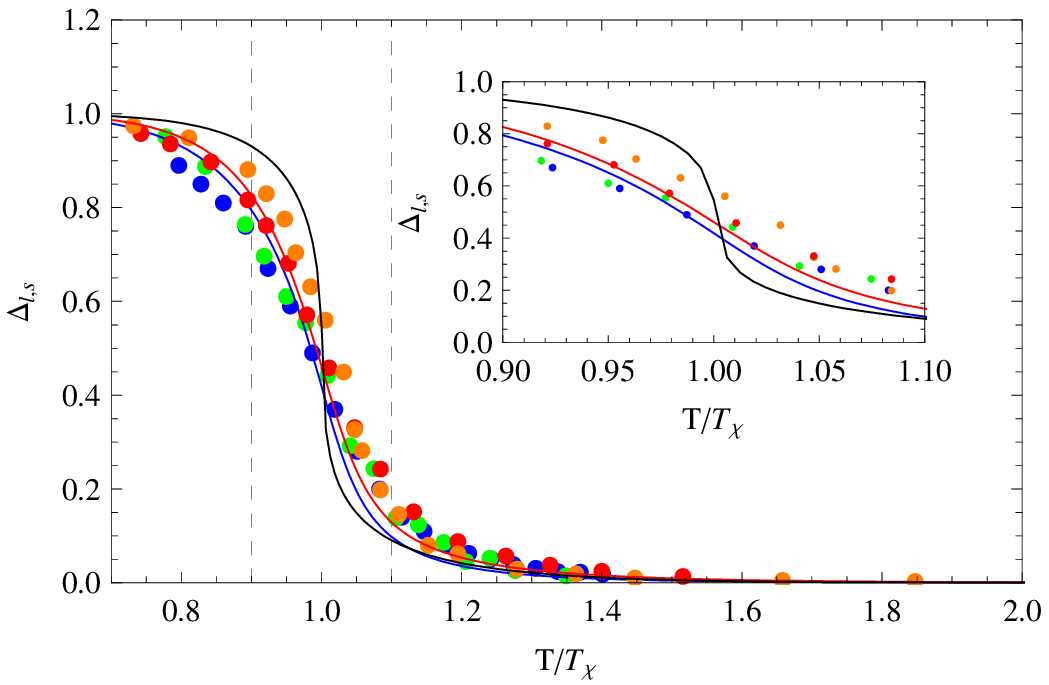}}
\end{center}
\caption{{\it Left}: Polyakov condensate $\langle\Phi\rangle$
obtained for ModelWB (solid blue) and ModelHotQCD (solid red) with the Polyakov potential
$\mathcal{U}_{\text{Poly-VM}}$. The dashed blue curve is $\langle\Phi\rangle$ obtained in ModelWB
with $\mathcal{U}_{\text{Poly}}$ as the Polyakov potential. For comparison we have also plotted 
$\langle\Phi\rangle$ obtained in PQM (solid black) with ModelHotQCD parameter set and $\mathcal{U}_{\text{Poly}}$
Polyakov potential. LQCD data from WB collaboration in blue (continuum estimate) and
green ($N_{\tau}=8$)~\cite{LQCD-WB1} and from HotQCD group in red ($N_{\tau}=8$ with asqtad action) and 
orange ($N_{\tau}=8$ with p4 action)~\cite{Lat-Baz} are also shown.
{\it Right}: Plots of $\Delta_{l,s}$ in ModelWB (blue) and ModelHotQCD (red) as obtained in PQMVT are shown.
The black curve is the model prediction in PQM with ModelHotQCD parameter set. All the model predictions are
obtained with the Polyakov potential $\mathcal{U}_{\text{Poly-VM}}$ . For comparison LQCD data are also shown: 
WB continuum estimate in blue~\cite{LQCD-WB1} and 
$N_{\tau}=10$~\cite{LQCD-WB2} data in green are shown, the HotQCD data for $N_{\tau}=8$ with p4 (orange) and 
asqtad (red) actions~\cite{Lat-Baz} are also shown.}
\label{fg.ordparam2}
\end{figure}

In the left panel of Fig.~\ref{fg.ordparam2}, we have plotted the Polyakov condensate $\langle\Phi\rangle$
obtained for both the parameter sets with $\mathcal{U}_{\text{Poly-VM}}$ as the Polyakov potential and compared 
with LQCD~\cite{LQCD-WB1,Lat-Baz}. We have also plotted $\langle\Phi\rangle$ with $\mathcal{U}_{\text{Poly}}$ as the Polyakov potential and
ModelWB as the parameter set for comparison. Although the VanderMonde term in $\mathcal{U}_{\text{Poly-VM}}$ suppresses 
the $\langle\Phi\rangle$ value, the model prediction is still much higher as compared to lattice. The
use of a potential that yields first order phase transition in case of pure glue theory results in a more
rapid rise of the Polyakov condensate not seen in lattice as also reported in PQM studies~\cite{PQMlat}.

A direct comparison of $\langle\sigma_x\rangle$ and $\langle\sigma_y\rangle$ with lattice data is not possible. 
Unknown normalisaton factors need to be first removed. One such observable
with the desired chiral limit and which also acts as an order parameter for chiral symmetry breaking
is~\cite{Lat-Cheng,Lat-Baz}
\beq
 \Delta_{l,s}\l T\r=\frac{\langle\bar{\psi}\psi\rangle_{l,T}-
 \frac{\hat{m_l}}{\hat{m_s}}\langle\bar{\psi}\psi\rangle_{s,T}}
 {\langle\bar{\psi}\psi\rangle_{l,0}-
 \frac{\hat{m_l}}{\hat{m_s}}\langle\bar{\psi}\psi\rangle_{s,0}}
\label{deltalslat}
\eeq
In our model computation, $\Delta_{l,s}$ will become~\cite{PQMlat}
\beq
 \Delta_{l,s}\l T\r=\frac{\langle\sigma_x\rangle\l T\r-
 \l\frac{h_x}{h_y}\r\langle\sigma_y\rangle\l T\r}
 {\langle\sigma_x\rangle\l0\r-
 \l\frac{h_x}{h_y}\r\langle\sigma_y\rangle\l0\r}
\label{deltalsmod}
\eeq

In the right panel of Fig.~\ref{fg.ordparam2}, we present the comparison of $\Delta_{l,s}$ obtained in PQMVT 
with LQCD~\cite{LQCD-WB1,LQCD-WB2,Lat-Baz} for both the parameter sets. We also compare with results from PQM. A similar 
analysis made in~\cite{PQMlat} concluded 
that the transition of $\Delta_{l,s}$ is sharper in PQM than that obtained in lattice. Here we find that the 
inclusion of the vacuum term in PQMVT makes the drop in $\Delta_{l,s}$ across the crossover region much more gentle.
This leads to better agreement with lattice data. Thus the crossover transition becomes smoother with the addition 
of the vacuum term. We will see that the smooth variation of the condensates 
across the transition temperature also result in smooth variation of all thermodynamic quantities and has
important consequence on the phase diagram as well.

\subsection{\text{$p$, $\epsilon$, $\Delta$, $s$, $c_{v}$, $c_s^2$} }

\begin{figure}
 \begin{center}
  \scalebox{0.75}{\includegraphics{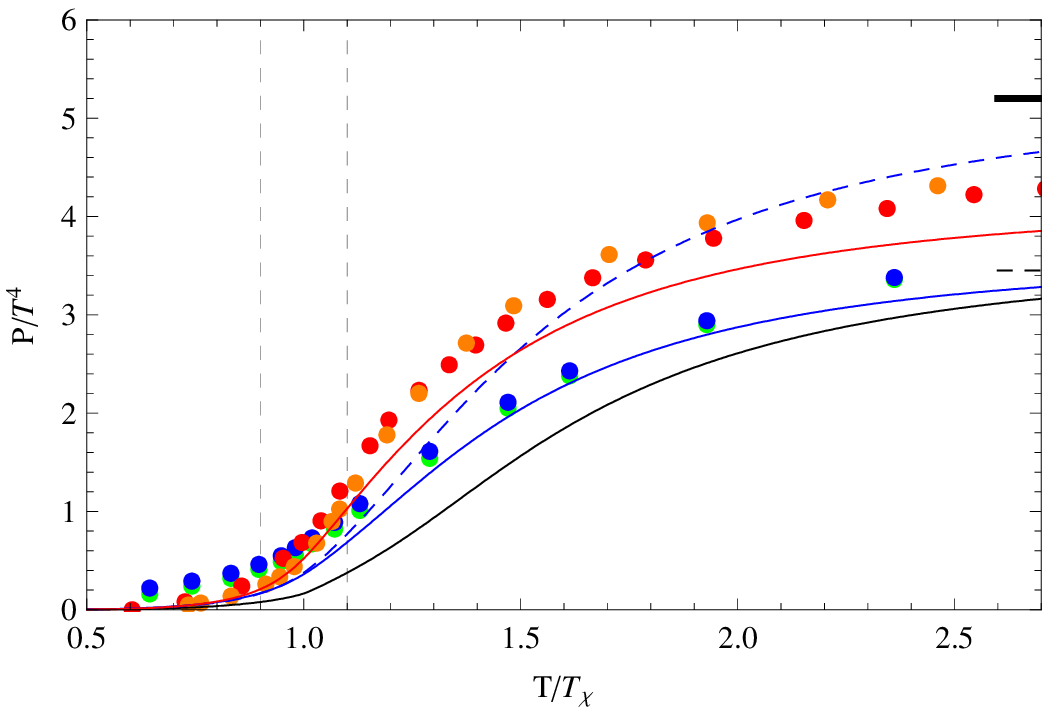}}
  \scalebox{0.75}{\includegraphics{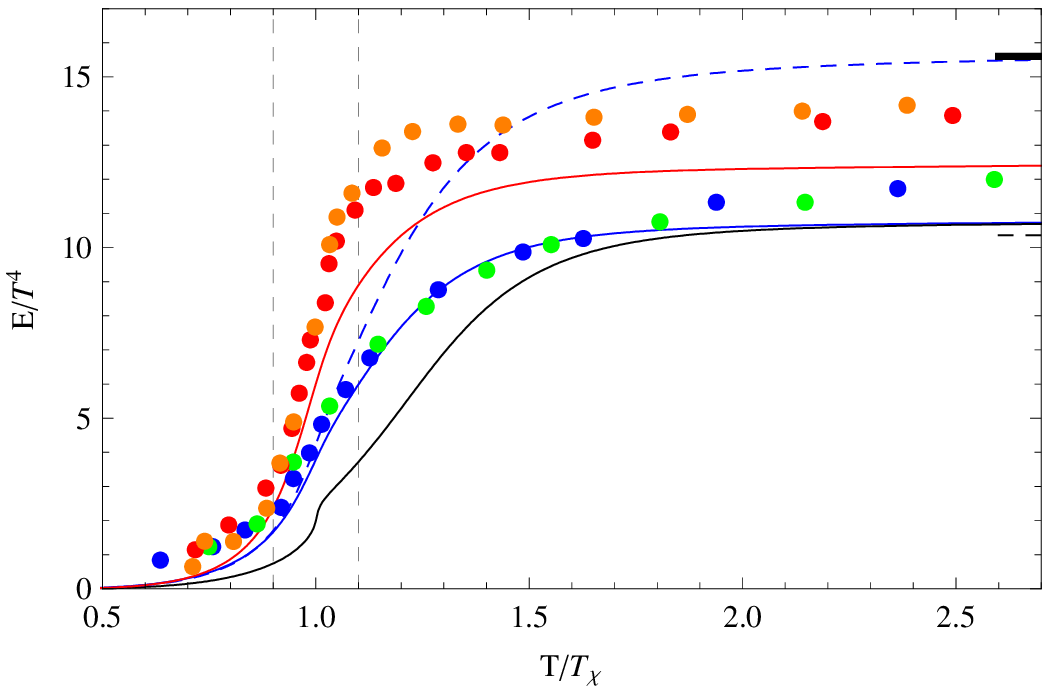}}\\
  \scalebox{0.75}{\includegraphics{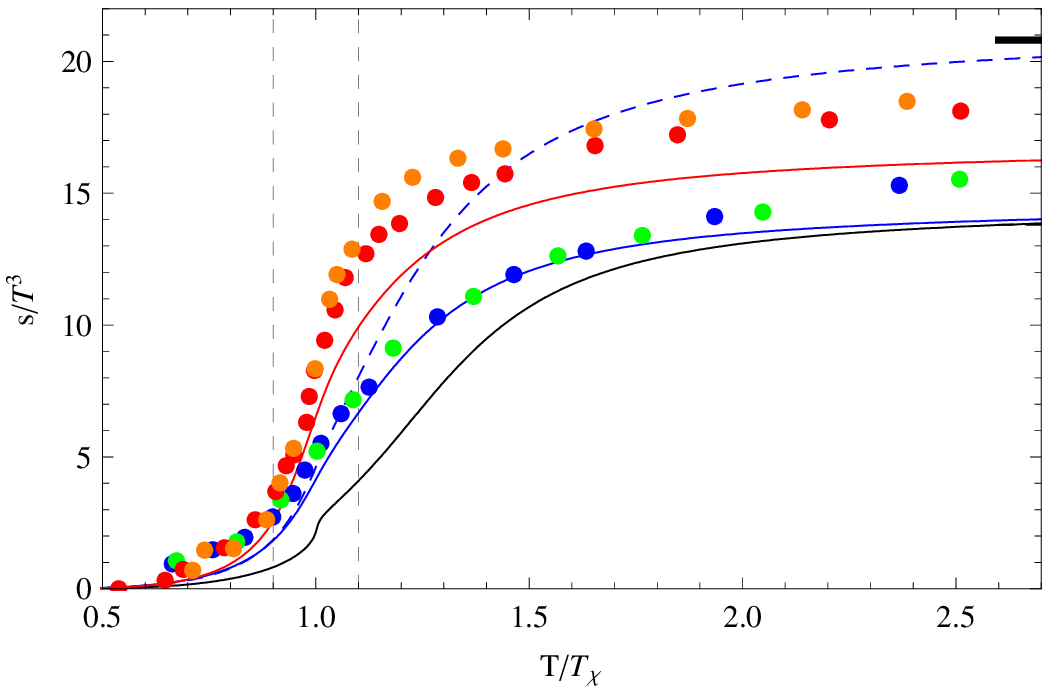}}
  \scalebox{0.75}{\includegraphics{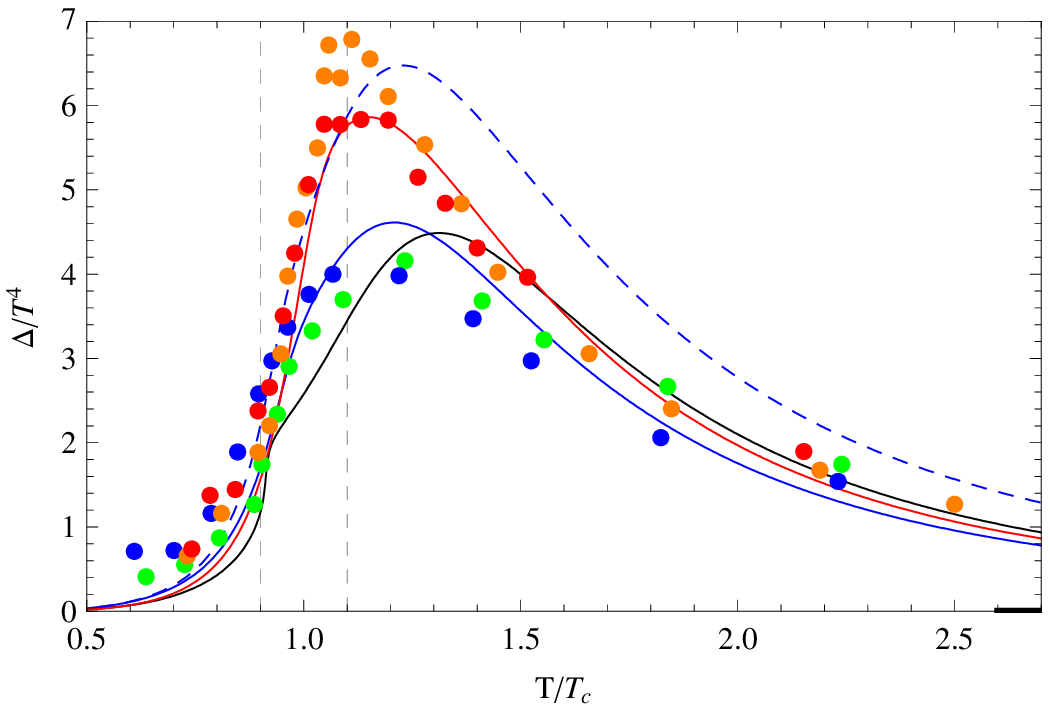}}
 \end{center}
\caption{Plots of $p$, $\epsilon$, $s$ and $\Delta$ in PQMVT with ModelWB (blue solid)
and ModelHotQCD (red solid) parameter sets. The corresponding black curves 
are the PQM predictions with ModelWB parameter set. These are obtained with the Polyakov potential 
$\mathcal{U}_{\text{Poly-VM}}$. For comparison, model predictions with the Polyakov potential 
$\mathcal{U}_{\text{Poly}}$ and ModelWB parameter set are also shown in dashed blue in each case. 
Here $T_c=\frac{T_{\chi}+T_d}{2}$. The LQCD data from WB are plotted in green and blue while
the HotQCD data are plotted in orange and red. In case of $p$, the continuum estimate in blue and
$N_{\tau}=10$ WB data in green~\cite{Lat-WB} are shown. Also shown are the $N_{\tau}=8$ (red)\cite{Lat-Baz}
and  $N_{\tau}=6$ (orange)\cite{Lat-Cheng} HotQCD data. For both $\epsilon$ and $s$, LQCD data from WB are shown for
$N_{\tau}=10$ (blue) and $N_{\tau}=8$ (green)~\cite{Lat-WB} and HotQCD $N_{\tau}=8$~\cite{Lat-Baz} (red)
and $N_{\tau}=6$~\cite{Lat-Cheng} data are also shown. For $\Delta$, WB data of continuum estimate~\cite{Lat-WB} in blue
and $N_{\tau}=10$~\cite{LQCD-WB2} in green are shown while HotQCD $N_{\tau}=8$ data with p4 (orange) and 
asqtad (red)~\cite{Lat-Baz} actions are also shown. In all the curves the high $T$ SB limit of
an ideal gas of massless fermions and gluons as in (\ref{eq.pSB}) is indicated by a thick black line 
while the dashed black line indicates the SB limit of an ideal gas of fermions only as in (\ref{eq.pSB})
without the first term. }
\label{fg.TQ}
\end{figure}

While on the lattice, the usual convention is to compute the trace of the energy momentum tensor~\cite{Lat-Baz}
and extract the rest of the thermodynamic quantities from it, in model calculations a different approach is
followed. Here it is natural to first compute the grand potential $\Omega$ with the values of the condensates
obtained by solving (\ref{eq.gapeq}). From $\Omega$ all the thermodynamic quantities can be easily obtained.
The pressure of the system is given by
\beq
p\l T,\mu_x,\mu_y\r=-\Omega\l T,\mu_x,\mu_y\r
\label{pressure}
\eeq 
with the vacuum normalization $p\l0,0,0\r=0$. The energy density $\epsilon$ is given by
\beq
\epsilon=-T^2\frac{\partial\l\Omega/T\r}{\partial T}=-T\l\frac{\partial\Omega}{\partial T}\r+\Omega
\label{e}
\eeq
while the entropy density $s$ is now easily obtained from the thermodynamic relation at zero chemical
potential
\beq
\epsilon=-p+Ts
\label{s}
\eeq
Using (\ref{pressure}) and (\ref{e}), we can find the conformal symmetry breaking measure $\Delta$ given
by the trace of the energy momentum tensor $\Theta^{\mu\mu}\l T\r$.
\beq
\frac{\Delta}{T^4}=\frac{\Theta^{\mu\mu}}{T^4}=\frac{\epsilon-3p}{T^4}
\eeq

In Fig. (\ref{fg.TQ}), we have plotted these thermodynamic quantities as obtained in PQMVT for both the
parameter sets, ModelWB and ModelHotQCD with $\mathcal{U}_{\text{Poly-VM}}$ as the Polyakov loop
potential. We have also plotted the PQM results with $\mathcal{U}_{\text{Poly-VM}}$ and PQMVT results
with $\mathcal{U}_{\text{Poly}}$ for comparison. We find good qualitative agreement overall with LQCD.
The smooth rise of $p$, $\epsilon$ and $s$ with $T$ for model as well as LQCD indicate 
the crossover nature of the phase transition that takes place at $\mu=0$~\cite{LQCD-crossover1,LQCD-crossover2,
LQCD-crossover3,LQCD-crossover4,PQMlat}. At low $T$, we see that $\Phi$ and $\bar{\Phi}$ $\simeq 0$ and so it is clear from
(\ref{eq.gf}) that most of the contribution to thermodynamic quantities like $p$, $\epsilon$ and $s$ come
from the three-quark ``hadron`` states. This is how statistical confinement works generically
in these model studies~\cite{stat-confine}. As is evident from Fig.~(\ref{fg.ordparam}), high
values of $\sigma_x$ and $\sigma_y$ result in large mass for these effective degrees of freedom
yielding suppressed values for all these thermodynamic quantities at lower temperatures.
At larger temperatures, the melting of the quark condensates across $T_{\chi}$ (see Fig.~(\ref{fg.ordparam})) 
result in lighter effective degrees of freedom. Also, $\Phi$ and $\bar{\Phi}$ $\rightarrow 1$ resulting
in the release of color states as the single quark states become the dominant degrees of freedom. This is 
manifested by all the three thermodynamic quantities that steadily rise across $T_{\chi}$ and saturate to 
a much higher value close to that of an ideal gas of $N_f$ massless fermions and $\l N_c^2-1\r$ gluons whose 
pressure $p_{SB}$ is given by
\beq
\frac{p_{SB}}{T^4}=\l N_c^2-1\r\frac{\pi^2}{45}+N_cN_f\frac{7 \pi^2}{180}.
\label{eq.pSB}
\eeq
There is good agreement between ModelWB predictions for $p$, $\epsilon$ and $s$ with the 
corresponding WB $N_{\tau}=8$, 10 and continuum estimate~\cite{Lat-WB} data in the temperature 
range $\approx 0.8 - 2$ $T_{\chi}$. At lower temperatures the agreement between model predictions
and LQCD is not good which could be due to the mean field treatment of the mesons.
Comparison of LQCD data with model predictions after taking into account mesonic fluctuations~\cite{PQMfluc}
should be carried out in order to throw more light on this issue. At higher temperatures the model description 
may not match well with LQCD as the contribution from the transverse gluons become significant and their
physics may not be captured by the Polyakov loop $\Phi$. The ModelHotQCD prediction although qualitatively
in agreement with LQCD, lies slightly below the HotQCD data for all temperatures. However, it should
be noted that the $N_{\tau}=8$ data~\cite{Lat-Baz} is consistently lower than that of $N_{\tau}=6$~\cite{Lat-Cheng} 
and so we expect that the agreement will improve when the continuum estimates will be available.

The model predictions for the conformal symmetry breaking measure $\Delta$ has also been plotted
and compared with LQCD. In this case we have plotted $\Delta$ against $T_c=\frac{T_{\chi}+T_d}{2}$ for better
agreement between model and LQCD. While for lower and higher temperatures, HotQCD~\cite{Lat-Baz} and WB~\cite{Lat-WB, LQCD-WB2} data look 
reasonably consistent, around $1 - 1.5$ $T_c$ although both exhibit a peak, the height of the peak in case
of HotQCD is almost $50 \%$ greater than that of WB. Moreover, the $N_{\tau} = 8$ HotQCD data with asqtad action is
significantly lower than that of p4 action at the peak. Thus the LQCD prediction in this temperature regime
is highly sensitive to the lattice volume as well as the lattice action implemented. As mentioned earlier,
we have chosen the value of $\kappa$ in (\ref{eq.VMpot}) such that the model prediction agrees well with LQCD.  In general 
$\kappa$ could be taken as a function of $T$ but in this work we have taken a constant value for $\kappa$.

The bumpy structures in $\langle\sigma_y\rangle$ and $\langle\Phi\rangle$ that we found around $T_{\chi}$ in 
Fig.~\ref{fg.ordparam} in the case of PQM shows up in the thermodynamic quantities also. Although the
pressure looks smooth across the transition region, others like the $\epsilon$, $s$ and $\Delta$
which are functions of $\frac{\partial\Omega}{\partial T}$ exhibit bumps around $T_{\chi}$. 
As observed in Fig.~(\ref{fg.ordparam}), addition of the vacuum term in PQMVT results in a smooth 
behaviour of the condensates in the transition region which in turn smoothens all the thermodynamic quantities in
Fig.~(\ref{fg.TQ}).

\begin{figure}
 \begin{center}
  \scalebox{0.75}{\includegraphics{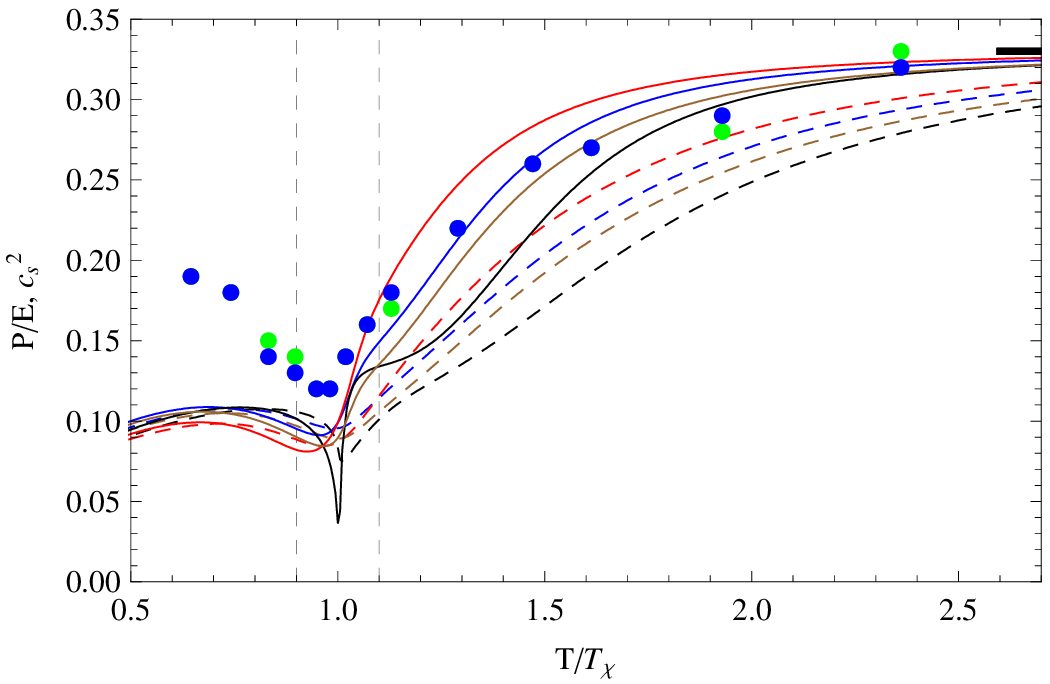}}
  \scalebox{0.75}{\includegraphics{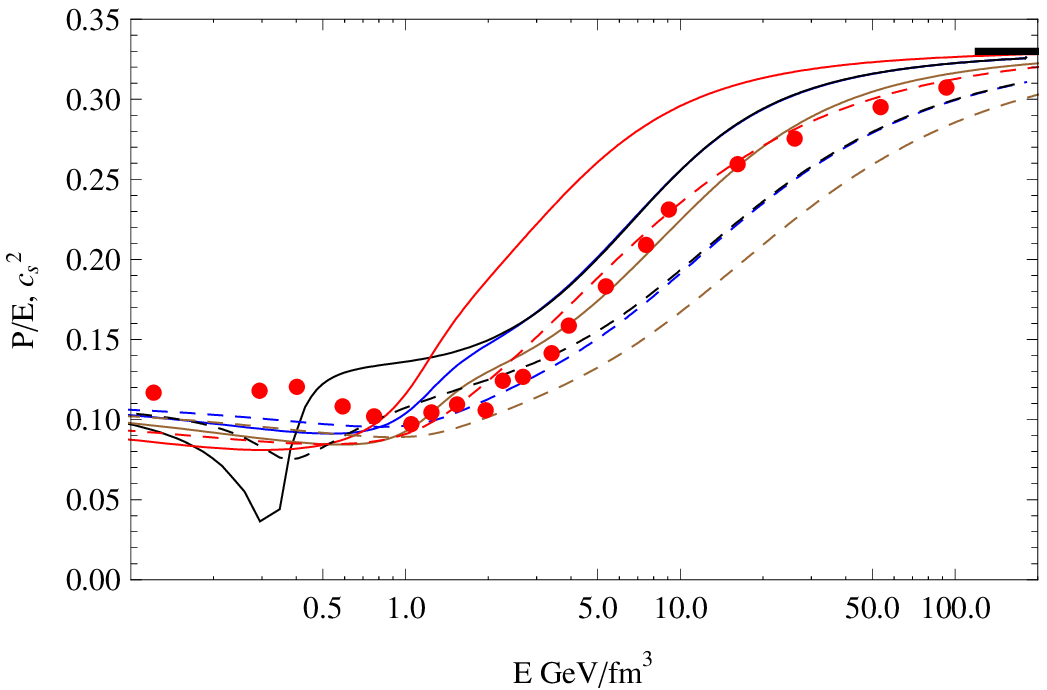}}
 \end{center}
\caption{Plots of  $c_s^2$ (solid) and $p/\epsilon$ (dashed) are shown in PQMVT with ModelWB (blue)
and ModelHotQCD (red) parameter sets. The corresponding black curves 
are the PQM predictions with ModelWB parameter set. These are obtained with the Polyakov potential 
$\mathcal{U}_{\text{Poly-VM}}$. Also shown are the plots with $\mathcal{U}_{\text{Poly}}$ with ModelWB 
parameter set in brown for comparison. WB data of $c_s^2$ for $N_{\tau}=10$ in green and the continuum estimate in 
blue~\cite{Lat-WB} are shown. Also shown are the HotQCD $N_{\tau}=8$ data~\cite{Lat-Baz} of $p/\epsilon$ in red. In 
all the curves the SB limit is indicated by a thick black line.}
\label{fg.cs2}
\end{figure}

\begin{figure}
 \begin{center}
  \scalebox{0.75}{\includegraphics{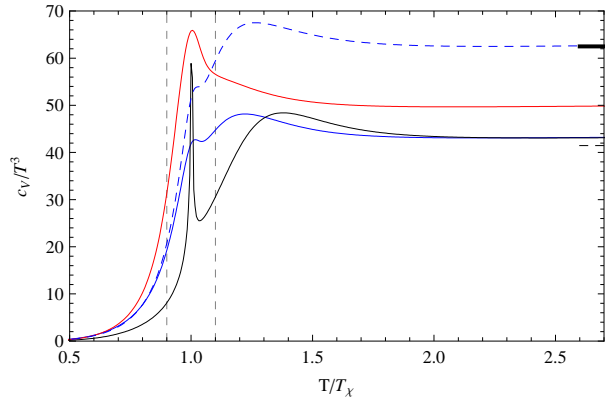}}
 \end{center}
\caption{Plots of  $c_V$ in PQMVT with ModelWB (blue) and ModelHotQCD (red) are shown. The black curve 
is the PQM prediction with ModelWB parameter set. These are obtained with the Polyakov potential 
$\mathcal{U}_{\text{Poly-VM}}$. Also shown is the plot with $\mathcal{U}_{\text{Poly}}$ with ModelWB 
parameter set (blue dashed) for comparison. The high $T$ SB limit of an ideal gas of
of massless fermions and gluons as in (\ref{eq.pSB}) is indicated by a thick black line 
while the dashed black line indicates the SB limit of an ideal gas of fermions only as in (\ref{eq.pSB})
without the first term.}
\label{fg.cv}
\end{figure}

Another important quantity from the point of view of hydrodynamical investigations of
relativistic heavy ion collisions is the isentropic speed of sound $c_s$ given by
\beq
 c_s^2=\left.\frac{\partial p}{\partial \epsilon}\right|_s=\left.\frac{\partial p}{\partial
 T}\right|_V\left/
\left.\frac{\partial \epsilon}{\partial T}\right|_V\right.=\frac{s}{c_V}
\eeq
where $c_V$ is the specific heat capacity at constant volume obtained as follows
\beq
c_V=\left.\frac{\partial \epsilon}{\partial T}\right|_V=\left.-T\frac{\partial^2\Omega}{\partial T^2}\right|_V
\eeq
$c_s^2$ is also closely related to the equation of state parameter $p/\epsilon$~\cite{Lat-Baz}
\beq
c_s^2=\frac{\partial p}{\partial \epsilon}=\epsilon\frac{\partial}{\partial \epsilon}\l\frac{p}{\epsilon}\r+\l\frac{p}{\epsilon}\r
\label{cs2-pbye}
\eeq

In Fig.~\ref{fg.cs2} we have plotted the model prediction for $c_s^2$ and $p/\epsilon$ against $T$ and 
also $\epsilon$. In the latter case the uncertainties due to $T_{\chi}$ go away.
In Fig.~\ref{fg.cv} we have plotted the model prediction for $c_v$.
The bumpy structures seen in Fig.~\ref{fg.TQ} for PQM becomes even more prominent in $c_s^2$ and $c_V$ 
which depend on $\frac{\partial^2\Omega}{\partial T^2}$. $c_V$ exhibits a sharp spike like structure at $T_{\chi}$ 
which results in a sharp dip in $c_s^2$ at the same temperature. The PQMVT plots are
lot smoother and compare well with LQCD qualitatively. Around $T_{\chi}$, WB continuum estimate for $c_s^2$~\cite{Lat-WB} show a minimum
with a value around 0.12 while ModelWB also exhibits a minimum in the same region with a minimum value $\sim 0.08$. 
In the high $T$ limit, both $c_s^2$ and $p/\epsilon$ for model and LQCD are found to approach the SB limit $1/3$.
In case of $c_V$ the high temperature limit is dependent on the Polyakov potential chosen. As seen in 
Fig.~(\ref{fg.cv}), the model prediction with $\mathcal{U}_{\text{Poly}}$ saturates to the SB limit
by $2 T_{\chi}$ whereas the model prediction with $\mathcal{U}_{\text{Poly-VM}}$ saturates to a lower value $\sim 73 \%$ of
the SB limit. Similarly even in the case of $p$, $\epsilon$ and $s$ in the high temperature we find the
model predictions with $\mathcal{U}_{\text{Poly-VM}}$ always lower than those with $\mathcal{U}_{\text{Poly}}$.

We end this subsection with an interesting observation. We find that the model prediction for the thermodynamic
quantities like $p$, $\epsilon$, $s$ and $c_V$ with ModelWB parameter set and  $\mathcal{U}_{\text{Poly-VM}}$ 
Polyakov potential for high temperatures saturate close to the SB limit of an ideal gas with only massless 
fermions and no gluons which is the $p_{SB}$ from (\ref{eq.pSB}) after removing the first term due to gluons.
In~\cite{PQMlat} similar results were reported when the Fukushima ansatz~\cite{Fukupot} for the Polyakov potential 
was used. Thus we find that the high temperature behaviour for model predictions with ModelWB parameter set and
$\mathcal{U}_{\text{Poly-VM}}$ is similar to that of the Fukushima potential~\cite{Fukupot} which is based on a 
strong coupling analysis.

\subsection{Quark Number Susceptibilities}

Quark number susceptibilities (QNS) are defined as derivatives of pressure with respect to quark chemical
potentials
\beq
\chi^{uds}_{ijk}=\frac{\partial^{i+j+k}(p/T^4)}{\partial \l \mu_u/T\r^i\partial \l \mu_d/T\r^j\partial 
\l \mu_s/T\r^k}
\label{eq.qns}
\eeq
Linear combinations of these can be related to fluctuations and correlations of conserved charges 
like baryon number, electric charge and strangeness which have been proposed as possible observables 
to signal the QCD phase transition~\cite{fluc1,fluc2}.
These have been measured in LQCD~\cite{Lat-Chengsus,LQCD-WB3,Lat-Baz,Hotlatnew,LQCD-WB4,LQCD-WB1}. Here we show results on QNS upto second
order. A detailed analysis of the fluctuations and correlations
of conserved charges in PQMVT and their comparison with LQCD will be done elsewhere~\cite{QNS-PQMVT}.
\begin{figure}
 \begin{center}
  \scalebox{0.75}{\includegraphics{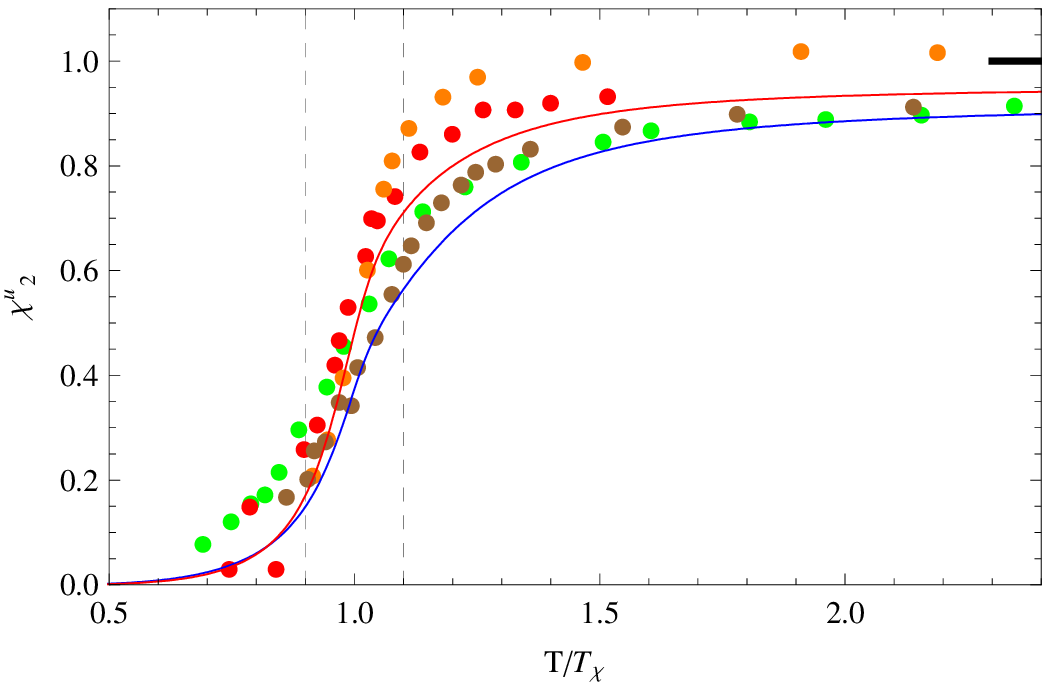}}
  \scalebox{0.75}{\includegraphics{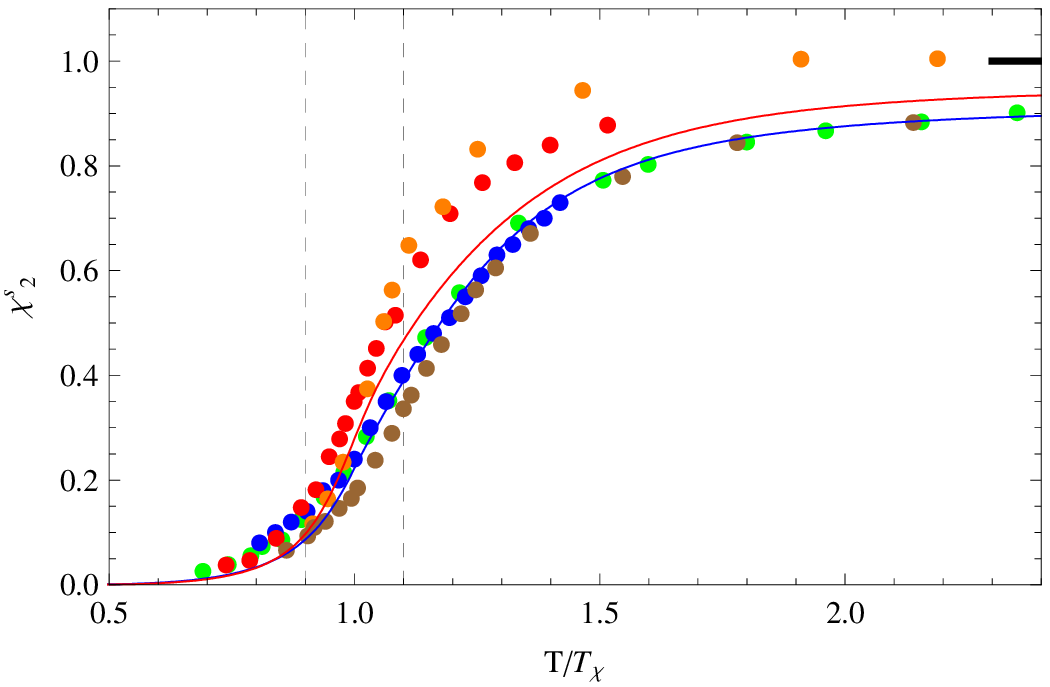}}\\
  \scalebox{0.75}{\includegraphics{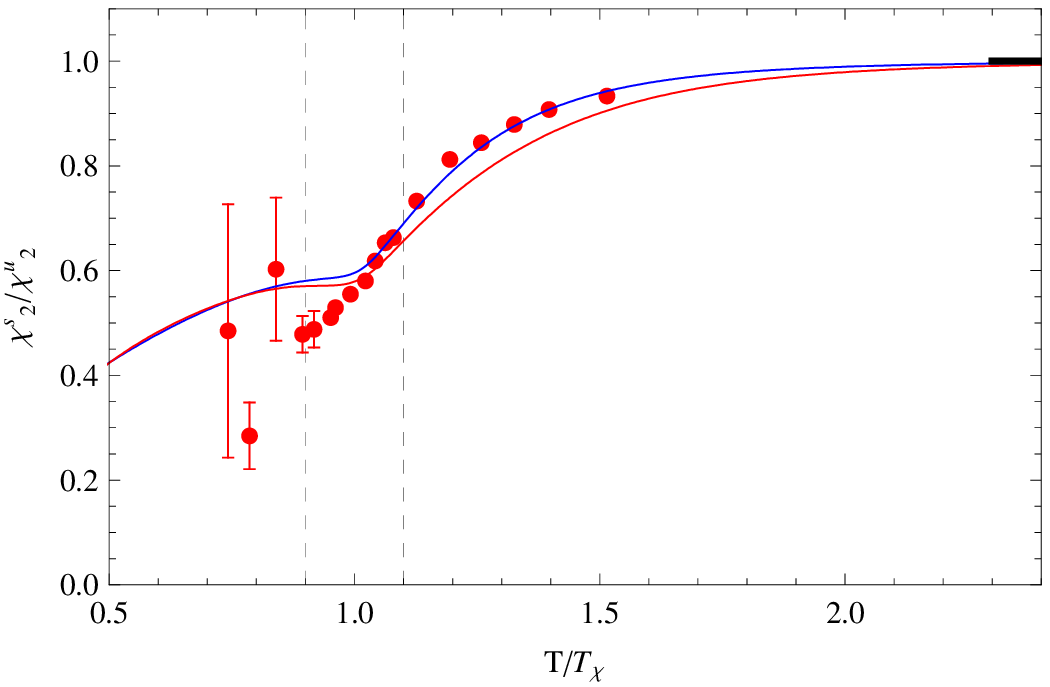}}
  \scalebox{0.75}{\includegraphics{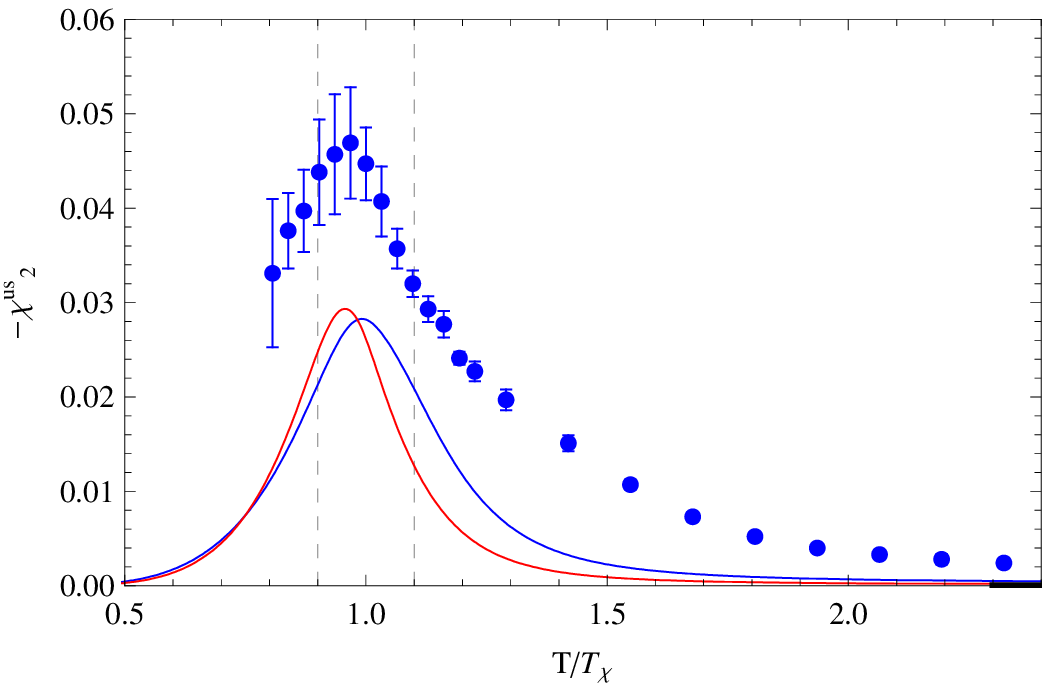}}
 \end{center}
\caption{QNS as obtained in PQMVT are plotted for ModelWB (blue) and ModelHotQCD (red) with the Polyakov potential 
$\mathcal{U}_{\text{Poly-VM}}$. For both $\chi_2^u$ and $\chi_2^s$, WB $N_{\tau}=12$ data~\cite{LQCD-WB3} (green) and
HotQCD $N_{\tau}=8$~\cite{Lat-Baz} (red), $N_{\tau}=6$~\cite{Lat-Chengsus} (orange) and $N_{\tau}=8$ with hisq 
action~\cite{Hotlatnew} (brown) are shown for comparison. Also in case of $\chi_2^s$, WB continuum estimate~\cite{LQCD-WB1} is
shown in blue. The HotQCD data for $\chi_2^s/\chi_2^u$ in red~\cite{Lat-Baz} and the WB continuum estimate for
$\chi_2^{us}$~\cite{LQCD-WB4} in blue are also compared with the model predictions. The high $T$ SB limit is indicated by a thick black line.}
\label{fg.sus}
\end{figure}

Fig. (\ref{fg.sus}) shows the diagonal susceptibilities  $\chi_2^u$ and $\chi_2^s$, their ratio $\chi_2^s/\chi_2^u$ and the 
off-diagonal $\chi_2^{us}$. All susceptibilities vanish at low temperature since the condensates have large values which
implies large masses of the relevant degrees of freedom that restricts the capacity of fluctuations.
At high temperature where
the degres of freedom have small masses and temperature is the only relevant scale, the system begins to behave like an
ideal gas of massless quarks. The susceptibilities thus approach the corresponding SB limit of ideal gas of three quarks.

The diagonal susceptibilities both show a monotonic increase with temperature, with the maximium rate of increase observed
around the crossover region and then saturating at higher temperatures to the SB limit. Both $\chi_2^u$ and $\chi_2^s$ saturate to 
about $80 - 90\%$ of the SB limit at $T\sim2.4 T_{\chi}$. We also observe that the light quark
susceptibility rises much faster than that of the strange quark both in model as well as LQCD. $\chi_2^s$ begins to 
saturate well beyond the cross-over region as is expected from the slow decrease of the value of the strange condensate.
For both $\chi_2^u$ and $\chi_2^s$ the $N_{\tau}=8$ data with asqtad action for HotQCD~\cite{Lat-Baz} lie consistently above ModelHotQCD.
On the other hand, there is much better agreement between WB data and ModelWB predictions. It is interesting to
observe that the recent $N_{\tau}=8$  susceptibility data with hisq action from HotQCD~\cite{Hotlatnew} which has also
been plotted in Fig. (\ref{fg.sus}) agree much better with the WB data.

It is useful to consider the ratio of the two susceptibilities, $\chi_2^s/\chi_2^u$, in order to avoid unknown normalization factors that
maybe present in lattice data \cite{Lat-Chengsus,QM}. In the unbroken phase there is excellent agreement between the models and lattice data, 
but they do not match as well below the crossover temperature. This maybe attributed to absence of mesonic fluctuations in a mean field 
approximation calculation \cite{PQMlat}. However, this ratio in the low temperature region is also highly dependent on the form of the 
Polyakov potential (see Fig.~(6 b) of \cite{PQMlat}). In our case the potential we use improves marginally the agreement of the model to
lattice predictions by giving smaller values to this ratio in the low temperature region. It should also be noted that the lattice
data in this region have large errors and comparisons of the model to lattice are uncertain.

We also show in Fig. (\ref{fg.sus}) the off-diagonal susceptibility $\chi_2^{us}$, which represents the correlation between the light and
heavy quarks. It vanishes in the SB limit, as is expected for a non-interacting ideal gas, and peaks in the crossover region.
Qualitatively the $\chi_2^{us}$ predictions of the model agree with lattice, showing the same essential features, however, lattice 
data is consistently larger over the entire temperature range. This maybe due to unknown normalization factors of lattice and 
perhaps a better quantity to compare would be normalized correlations \cite{QNS-PQMVT}. 

\subsection{Phase Diagram}

\begin{figure}
 \begin{center}
  \scalebox{0.95}{\includegraphics{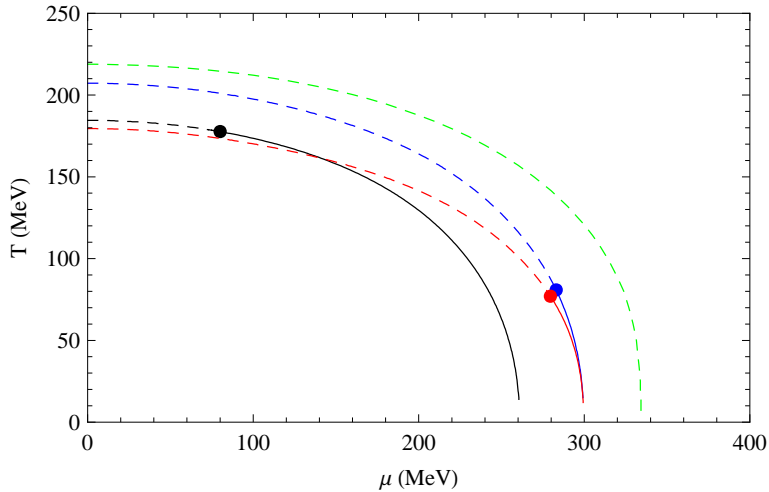}}
 \end{center}
\caption{ Phase diagrams for PQM (black) in ModelWB and PQMVT in ModelWB (blue) and ModelHotQCD (red) with the Polyakov potential 
$\mathcal{U}_{\text{Poly-VM}}$. The dashed lines indicate crossover, the solid lines indicate first order 
phase transition and the filled circles indicate CEP. The phase diagram for $m_{\sigma}=600$ MeV in the case of PQMVT with 
the Polyakov potential $\mathcal{U}_{\text{Poly}}$ is also shown (green).}
\label{fg.phdiag}
\end{figure}

The divergence of $\chi^u_2$ can be used to locate the CEP~\cite{CEP} where a first order phase
transition line ends in a second order phase transition point. We obtain the phase diagram in the
case of symmetric quark matter $\l\mu_x=\mu_y=\mu\r$. In Fig.~(\ref{fg.phdiag}) we
have presented the phase diagram for PQM and PQMVT. The chiral phase
boundary has been plotted and the CEP located in all the cases. The smoothening of the
thermodynamic quantities in the transition region as seen for $\mu=0$
persists even at non-zero chemical potentials. This results
in pushing the CEP to a higher value of $\mu$ in case of PQMVT as compared to PQM.  With $m_{\sigma}=400$ MeV, 
in PQM with ModelWB using the $\mathcal{U}_{\text{Poly-VM}}$ potential the CEP is found at $\l\mu,T\r=\l80,177.5\r$ MeV which gets shifted 
to $\l283,81\r$ MeV in case of PQMVT of ModelWB. We have also obtained the phase diagram in PQMVT with $m_{\sigma}=600$ MeV and
 $\mathcal{U}_{\text{Poly}}$ potential. In this case the 
phase diagram has no CEP and there is only crossover transition on the entire $\l\mu-T\r$ plane 
as in the case of 2 flavors~\cite{PQMVT-USG}.

We have also plotted the phase boundary of PQMVT in ModelHotQCD (red) with the $\mathcal{U}_{\text{Poly-VM}}$ potential. In 
this case $T_\chi$ is lower than that of ModelWB at $\mu=0$ that result in a lower phase boundary in the $\l T,\mu\r$ plane as compared
to the case with ModelWB parameter set.
This is the effect of choosing a smaller value of $T_0$. For large values of $\mu$ there is not much difference between the two phase boundaries.
This is essentially because in this region the Polyakov potential plays a less important role as compared to $\Omega_{q\bar{q}}$.
The phase boundary for ModelWB with $\mathcal{U}_{\text{Poly}}$  does not differ much from that of ModelWB with 
the $\mathcal{U}_{\text{Poly-VM}}$ potential implying that in this case the VanderMonde term does not play much of a role in the 
determination of the chiral phase boundary.

\section{Conclusion}
\label{sec.conc}

In this paper we have studied the effect of including the commonly neglected fermionic vacuum fluctuations to the $\l2+1\r$ PQM model.
The conventional PQM model suffers from a rapid phase transition contrary to what is found through lattice 
simulations~\cite{PQMlat}. This could be due to the use of a Polyakov potential which carries with it a 
remnant first order phase transition of the pure glue theory. Another possible reason could be the sudden 
release of quark degrees of freedom in PQM with increase in temperature. Addition of the vacuum 
term in PQMVT addresses the latter. This tames the rapid transitions that we see in the PQM model and 
significantly improves the model's agreement to lattice data.

We work with two Polyakov loop potentials: $\mathcal{U}_{\text{Poly}}$ and  $\mathcal{U}_{\text{Poly-VM}}$
given in (\ref{eq.polpot}) and (\ref{eq.VMpot}) respectively. The model parameters are fixed by inputs from
experiments as well as LQCD data. In this work, we have compared model predictions to LQCD data of HotQCD
and WB groups. Depending on the lattice group whose data we choose to fix the model parameters, we arrive
at the two sets of model parameters namely ModelWB and ModelHotQCD given in Table~\ref{tb.param}. We then
investigated the bulk thermodynamic properties of the system and the influence of the vacuum term in PQMVT. 
In PQM, $\Delta_{l,s}$ which is a suitably defined order parameter for chiral phase transition is 
found to undergo a much rapid fall across the transition temperatures compared to LQCD~\cite{PQMlat}. The inclusion
of the fermionic vacuum fluctuations in the model makes the transition much more gentle. The vacuum term essentially
washes away all the jagged and bumpy structures that are there in the quark codensates, Polyakov loop and all
thermodynamic quantities in PQM. We also computed the QNS upto second order. In all cases very good qualitative 
agreement between LQCD and model is found. The model computations suggest that the high temperature behaviour 
of the thermodynamic quantities with $\mathcal{U}_{\text{Poly-VM}}$ Polyakov potential and ModelWB parameter set is similar to that
of the Fukushima potential~\cite{Fukupot} for the Polyakov loop. We further investigated the role of the vacuum term 
on the phase diagram. The smoothening effect of the vacuum term persists even at non zero $\mu$. With $m_{\sigma}=600$ MeV,
we find that the vacuum term washes away the critical behaviour completely from the $\l\mu,T\r$ plane. In case
of $m_{\sigma}=400$ MeV, addition of the vacuum term pushes the CEP to larger chemical potential.

\section{Acknowledgement}
S.C. would like to acknowledge discussions and collaborations with Sourendu Gupta for introduction to 
the subject itself. He would also like to acknowledge many helpful discussions with Uma Shankar Gupta, 
Anirban Lahiri, Tanumoy Mandal, Apoorva Patel and Rajarshi Ray. We would like to thank Rohini Godbole 
for guidance. K.A.M. acknowledges the financial support through Junior Research Fellowship provided by CSIR.

\section{Appendix}
\label{sec.appendix}
\appendix
\section{Model Parameters}
\label{modelparameters}
The model parameters are determined by establishing the vacuum properties. In analogy to ~\cite{lenaghan,Schaefer:09}
the values of the condensates are determined from the pion and kaon decay constants by
means of the partially conserved axial-vector current relation (PCAC).
In the strange--non-strange basis they are given by
\begin{equation}
  \label{eq:pcac}
  \bsig_{x} = f_{\pi} \  ; \qquad
  \bsig_{y} = \frac{1}{\sqrt{2}} \left(2 \fk - f_{\pi}\right)\ .
%\label{condensates}
\end{equation} 
The remaining model parameters ($m^2$, $\lambda_1$, $\lambda_2$, $c$, $h_x$ and $h_y$) are fixed by the 
following inputs from experiments: pion mass $m_{\pi}$ and pion decay constant $f_\pi$, kaon mass $m_K$ 
and kaon decay constant $f_K$, average squared mass of $\eta$ and $\eta'$ mesons, $\l m_{\eta}^2+m_{\eta'}^2\r$ 
and sigma mass $m_\sigma$. In this appendix we shall derive the equations to determine the above parameters in PQMVT.

Using (\ref{eq.omega}) and (\ref{omegaqq}), we rewrite $\Omega\l T,\mu_x,\mu_y\r$ as
\beq
\Omega^{\text{PQMVT}}\l T,\mu_x,\mu_y\r=\Omega^{\text{PQM}}\l T,\mu_x,\mu_y\r+\Omega^v_{\bar{q}q}
\eeq
where 
\beq
\Omega^{\text{PQM}}\l T,\mu_x,\mu_y\r= U(\sigma_x, \sigma_y) + {\cal U_{\text{P}}} \left(\Phi,\bar{\Phi},
 T \right) +\Omega^{th}_{\bar{q}q}
\eeq
Now the meson masses are determined by the curvature of $\Omega$ at the global minimum
\beqa
 m_{\alpha,ab}^{2} &=& \frac{\partial^2 \Omega}{\partial \xi_{\alpha,a} 
 \partial \xi_{\alpha,b}} \bigg|_{\text{min}}\nn\\
&=&\frac{\partial^2 \Omega^{\text{PQM}}}{\partial \xi_{\alpha,a} 
 \partial \xi_{\alpha,b}} \bigg|_{\text{min}}+\frac{\partial^2 \Omega^{\text{v}}_{\bar{q}q}}{\partial
 \xi_{\alpha,a}\partial \xi_{\alpha,b}} \bigg|_{\text{min}}\nn\\
 &=&{m^{\text{PQM}}_{\alpha,ab}}^{2}+{m^{\text{v}}_{\alpha,ab}}^{2}
\label{eq.mass}
\eeqa
${m^{\text{PQM}}_{\alpha,ab}}^{2}$ are available in the literature~\cite{Schaefer:09}. Here we compute 
${m^{\text{v}}_{\alpha,ab}}^{2}$. From (\ref{eq.omegareg}) and (\ref{eq.mass}) we get
\beq
{m^{\text{v}}_{\alpha,ab}}^{2}=-\frac{N_c}{8\pi^2}\sum_f\left[\l
 2\log\l\frac{m_f}{\Lambda}\r+\frac{3}{2} \r\l\frac{\partial m_f^2}{\partial
 \xi_{\alpha,a}^2}\r\l\frac{\partial m_f^2}{\partial \xi_{\alpha,b}^2}\r+\l
 \frac{m_f^2}{2}+2m_f^2\log\l\frac{m_f}{\Lambda} \r \r \frac{\partial^2
  m_f^2}{\partial \xi_{\alpha,a}\partial \xi_{\alpha,b}}\right]
\eeq
The expressions for $\l\frac{\partial m_f^2}{\partial
 \xi_{\alpha,a}}\r\l\frac{\partial m_f^2}{\partial \xi_{\alpha,b}}\r$ and $\frac{\partial^2
  m_f^2}{\partial \xi_{\alpha,a}\partial \xi_{\alpha,b}}$ are already available in~\cite{Schaefer:09}.
In Table~(\ref{tb.masses}), we present the expressions of $m^2_{\alpha,ab}$ for the different mesons
as obtained in PQMVT at $T=\mu=0$.

\begin{table}
 \begin{center}
\begin{tabular}{|l|l|} 
\hline \hline
Mesons&Expression for masses\\ \hline
$ m^{2}_{\pi}$ & $m^2 + \lambda_1 (x^2 + y^2) +\frac{\lambda_2}{2} x^2 -\frac{\sqrt{2} c}{2} y
-\frac{N_cg^4}{64\pi^2}x^2X$  \\ %\hline
$m^{2}_{K}$ &$m^2 + \lambda_1 (x^2 + y^2) +\frac{\lambda_2}{2} (x^2 - \sqrt{2} x y +2 y^2) - \frac{c}{2} x
-\frac{N_cg^4}{64\pi^2}\l\frac{x-\sqrt{2}y}{x^2-2y^2}\r \l x^3X+2\sqrt{2}y^3Y\r$  \\ 
$m^2_{p,00}$ & $m^2 + \lambda_1(x^2 +y^2) + \frac{\lambda_2}{3}(x^2 +y^2) + \frac{c}{3} (2x + \sqrt{2} y)
-\frac{N_cg^4}{96\pi^2}\l x^2X+y^2Y\r$ \\
$m^2_{p,88}$ & $m^2 +\lambda_1(x^2 +y^2) +\frac{\lambda_2}{6}(x^2 +4y^2)-\frac{c}{6}(4x -\sqrt{2}y)
-\frac{N_cg^4}{192\pi^2}\l x^2X+4y^2Y\r$ \\
$\l m^{2}_{\eta}+m^{2}_{\eta'}\r$&$m^2_{p,00}+m^2_{p,88}$\\
$m^2_{s,00}$ &  $m^2 + \frac{\lambda_1}{3} (7 x^2 + 4 \sqrt{2} xy + 5 y^2) + \lambda_2(x^2 + y^2) - \frac{\sqrt{2}c}{3} (\sqrt{2} x +y)
-\frac{N_cg^4}{96\pi^2}\l3\l x^2X+y^2Y\r+4\l x^2+y^2\r\r$ \\
$m^2_{s,88}$ &  $m^2 + \frac{\lambda_1}{3} (5x^2 -4 \sqrt{2} xy +7y^2) + \lambda_2(\frac{x^2}{2} +2y^2) + \frac{\sqrt{2}c}{3} (\sqrt{2}x - \frac{y}{2})
-\frac{N_cg^4}{96\pi^2}\l \frac{3}{2}\l x^2X+4y^2Y\r+2\l x^2+4y^2\r\r$\\
$m^2_{s,08}$ &  $\frac{2\lambda_1}{3}(\sqrt{2}x^2 -xy -\sqrt{2}y^2) +\sqrt{2}\lambda_2(\frac{x^2}{2}-y^2) +\frac{c}{3\sqrt{2}}(x- \sqrt{2}y)$
-$\frac{N_cg^4}{8\sqrt{2}\pi^2}\l\frac{1}{4}\l x^2X-2y^2Y\r+\frac{1}{3}\l x^2-2y^2\r\r$\\
$ m^{2}_{\sigma} $& $m^2_{s,{00}} \cos^{2}\theta_s + m^2_{s,88}\sin^{2}\theta_s + 2  m^{2}_{s,08} \sin \theta_s \cos\theta_s$, where $\tan\l2\theta_s\r=\frac{2m^2_{s,08}}{m^2_{s,00}-m^2_{s,88}}$\\
  \hline
\end{tabular}
\end{center}
\caption{The expressions of squared meson masses in the vacuum required to determine the model parameters. $x$ denotes 
$\sigma_x$, $y$ denotes $\sigma_y$, $X$ denotes $\l1+4\log\l \frac{g\sigma_x}{2\Lambda}\r\r$ and $Y$ 
denotes $\l1+4\log\l \frac{g\sigma_y}{\sqrt{2}\Lambda}\r\r$.}
\label{tb.masses}
\end{table}
From the expressions of $m^2_{\pi}$, $m^2_K$ and $\l m^{2}_{\eta}+m^{2}_{\eta'}\r$ we find $\lambda_2$ and $c$
\beqa
\lambda_2&=&\frac{3\l2f_K-f_{\pi}\r m^{'2}_K-\l2f_K+f_{\pi}\r m^{'2}_{\pi}-2\l m^{2}_{\eta}+m^{2}_{\eta'}\r'
\l f_K-f_{\pi}\r}{\l3f^2_{\pi}+8f_K\l f_K-f_{\pi}\r\r\l f_K-f_{\pi}\r}\nn\\
c&=&\frac{m^{'2}_K-m^{'2}_{\pi}}{f_K-f_{\pi}}-\lambda_2\l2f_K-f_{\pi}\r
\label{c}
\eeqa
where
\beqa
m^{'2}_K&=&m^{2}_K+\frac{N_cg^4}{64\pi^2}\l\frac{x-\sqrt{2}y}{x^2-2y^2}\r \l x^3X+2\sqrt{2}y^3Y\r\nn\\
m^{'2}_{\pi}&=&m^{2}_{\pi}+\frac{N_cg^4}{64\pi^2}x^2X\nn\\
\l m^{2}_{\eta}+m^{2}_{\eta'}\r'&=&\l m^{2}_{\eta}+m^{2}_{\eta'}\r+\frac{N_cg^4}{96\pi^2}\l x^2X+y^2Y\r+
\frac{N_cg^4}{192\pi^2}\l x^2X+4y^2Y\r
\eeqa
Here $x$ denotes $\sigma_x$, $y$ denotes $\sigma_y$, $X$ denotes $\l1+4\log\l \frac{g\sigma_x}{2\Lambda}\r\r$ and 
$Y$ denotes $\l1+4\log\l \frac{g\sigma_y}{\sqrt{2}\Lambda}\r\r$.
Having obtained $\lambda_2$ and $c$, we use the expression of $m^2_{\pi}$ to express $m^2$ in terms of $\lambda_1$
\beq
m^2=m_{\pi}^{'2}-\frac{\lambda_2}{2}f^2_{\pi}+\frac{c}{2}\l2f_K-f_{\pi}\r-\lambda_1\l f^2_{\pi}+\frac{\l2f_K-f_{\pi}\r^2}{2}\r
\eeq 
Finally, $\lambda_1$ may be obtained from the expression of $m^2_{\sigma}$ while the explicit symmtery breaking 
parameters $h_x$ and $h_y$ are obtained from the stationarity conditions (\ref{eq.gapeq})
\beqa
 h_x&=&f_{\pi}m^{'2}_{\pi}-\frac{N_cg^4\sigma_x^3}{64\pi^2}\l1+4\log\l\frac{g\sigma_x}{2\Lambda}\r\r\\
h_y&=&\sqrt{2}f_Km_K^{'2}-\frac{f_{\pi}m^{'2}_{\pi}}{\sqrt{2}}-\frac{N_cg^4\sigma_y^3}{32\pi^2}
\l1+4\log\l\frac{g\sigma_y}{\sqrt{2}\Lambda}\r\r
\eeqa

Thus the final expression of the parameters are
\beqa
 h_{x} &=&f_{\pi} m^2_{\pi} ;\nn\\
 h_{y} &=&\frac{2 f_K m_K^2-f_{\pi } m_{\pi }^2}{\sqrt{2}} ; \nn\\
 \lambda_2\l\Lambda\r &=&\left(g^4 \text{Log}\left[\frac{g f_{\pi }}{2 \Lambda }\right] f_{\pi }^2
\left(-8 f_K^2+8 f_K f_{\pi }-3 f_{\pi }^2\right) N_c\right.+\nn\\
   &&g^4 \text{Log}\left[\frac{g \left(2 f_K-f_{\pi }\right)}{2 \Lambda }\right] \left(-2 f_K+f_{\pi }
\right){}^2 \left(8 f_K^2-8 f_K f_{\pi }+3 f_{\pi }^2\right) N_c+\nn\\
   &&f_K \left(8 g^4 f_K^3 N_c-16 g^4 f_K^2 f_{\pi } N_c-f_{\pi } \left(32 \pi ^2 \left(3 m_K^2+m_{\pi }^2
-2 \left(m_{\eta }^2+m_{\eta '}^2\right)\right)+3 g^4 f_{\pi }^2 N_c\right)+\right. \nn\\
   &&\left.\left.f_K \left(-64 \pi ^2 \left(-3 m_K^2+m_{\pi }^2+m_{\eta }^2+m_{\eta '}^2\right)+
11 g^4 f_{\pi }^2 N_c\right)\right)\right)/\left(32 \pi ^2 f_K \left(8 f_K^3-16 f_K^2 f_{\pi }+11 f_K 
f_{\pi }^2-3 f_{\pi }^3\right)\right) ;\nn\\
 c &=& -\frac{2 \left(2 f_K \left(m_K^2+m_{\pi }^2-m_{\eta }^2-m_{\eta '}^2\right)+f_{\pi } \left(-2
       m_{\pi }^2+m_{\eta }^2+m_{\eta '}^2\right)\right)}{8 f_K^2-8 f_K f_{\pi }+3 f_{\pi }^2} ;\nn\\
 m^2&=&\left(-g^4 \text{Log}\left[\frac{ \left(2 f_K-f_{\pi }\right)}{f_{\pi}}\right] f_{\pi }^2
       \left(-2 f_K+f_{\pi }\right){}^2 \left(8 f_K^2-8 f_K f_{\pi }+3 f_{\pi }^2\right)
        N_c-\right.\nn\\
    &&32 \pi ^2 f_K \left(32 f_K^5 \lambda _1-96 f_K^4 f_{\pi } \lambda _1+f_{\pi }^3
      \left(-3 m_K^2+m_{\pi }^2+4 \left(m_{\eta }^2+m_{\eta '}^2\right)-9 f_{\pi }^2 \lambda
     _1\right)-\right.\nn\\
    &&4 f_K^2 f_{\pi } \left(3 m_K^2-3 m_{\pi }^2-4 \left(m_{\eta }^2+m_{\eta '}^2\right)+26 f_{\pi
      }^2 \lambda _1\right)+4 f_K^3 \left(-2 \left(-m_K^2+m_{\pi }^2+m_{\eta }^2+m_{\eta
     '}^2\right)+33 f_{\pi }^2 \lambda _1\right)+\nn\\
    &&\left.\left.f_K f_{\pi }^2 \left(10 m_K^2-8 m_{\pi }^2-12 \left(m_{\eta }^2+m_{\eta '}^2\right)+
45 f_{\pi }^2 \lambda _1\right)\right)\right)/\left(64 \pi ^2 f_K \left(8 f_K^3-16 f_K^2 f_{\pi }+
11 f_K f_{\pi }^2-3 f_{\pi }^3\right)\right) ;\nn\\
 m_{\sigma}^2&=&m^2_{s,00}\cos^2\theta_s+m^2_{s,88}\sin^2\theta_s+
               2m^2_{s,08}\sin\theta_s\cos\theta_s, \quad \text{where} \nn\\
m_{s,00}^2&=&\left(3 g^4 \text{Log}\left[\frac{\left(2 f_K-f_{\pi }\right)}{f_{\pi} }\right]
    f_{\pi }^2 \left(-2 f_K+f_{\pi }\right){}^2 \left(8 f_K^2-8 f_K f_{\pi }+3 f_{\pi
      }^2\right) N_c-\right.\nn\\
    &&2 f_K \left(32 f_K^5 \left(g^4 N_c-16 \pi ^2 \lambda _1\right)+f_K^4 f_{\pi } \left(-96 g^4 N_c+512 \pi ^2
 \lambda _1\right)+\right.\nn\\
   &&f_K f_{\pi }^2 \left(16 \pi ^2 
\left(-26 m_K^2-9 m_{\pi }^2+m_{\eta }^2+m_{\eta '}^2\right)+f_{\pi }^2 \left(45 g^4 N_c+16 \pi ^2
 \lambda _1\right)\right)-\nn\\
   &&8 f_K^2 f_{\pi } \left(16 \pi ^2 \left(-5 m_K^2-3 m_{\pi }^2+m_{\eta }^2+m_{\eta '}^2\right)+
f_{\pi }^2 \left(13 g^4 N_c+32 \pi ^2 \lambda _1\right)\right)+\nn\\
    &&f_{\pi }^3 \left(-16 \pi ^2 \left(-9 m_K^2-2 m_{\pi }^2+m_{\eta }^2+m_{\eta '}^2\right)+
f_{\pi }^2 \left(-9 g^4 N_c+48 \pi ^2 \lambda _1\right)\right)+\nn\\
    &&\left.\left.4 f_K^3 \left(32 \pi ^2 \left(-4 m_K^2-m_{\pi }^2+m_{\eta }^2+m_{\eta '}^2\right)+
f_{\pi }^2 \left(33 g^4 N_c+48 \pi ^2 \lambda _1\right)\right)\right)\right)/\left(96 \pi ^2 f_K
 \left(8 f_K^3-16 f_K^2 f_{\pi }+\right.\right.\nn\\
  &&\left.\left.11 f_K f_{\pi }^2-3 f_{\pi }^3\right)\right) ;\nn\\
m_{s,88}^2&=&\left(3 g^4 \text{Log}\left[\frac{ \left(2 f_K-f_{\pi }\right)}{f_{\pi} }\right]
             f_{\pi }^2 \left(-2 f_K+f_{\pi }\right){}^2 \left(8 f_K^2-8 f_K f_{\pi }+3 f_{\pi
             }^2\right) N_c-\right.\nn\\
          &&2 f_K \left(-64 f_K^4
          f_{\pi } \left(3 g^4 N_c-64 \pi ^2 \lambda _1\right)+64 f_K^5 \left(g^4 N_c-16 \pi ^2 \lambda _1\right)+
          \right.\nn\\&&f_K f_{\pi }^2 \left(16
          \pi ^2 \left(-58 m_K^2-39 m_{\pi }^2+32 \left(m_{\eta }^2+m_{\eta
         '}^2\right)\right)+f_{\pi }^2\left(57 g^4 N_c-2176 \pi ^2 \lambda _1\right)\right)-\nn\\
        &&32 f_K^2 f_{\pi } \left(4 \pi ^2 \left(-13 m_K^2-3 m_{\pi }^2+5 \left(m_{\eta }^2+m_{\eta
        '}^2\right)\right)+f_{\pi }^2 \left(5 g^4 N_c-164 \pi ^2 \lambda _1\right)\right)+\nn\\
        &&16 f_K^3 \left(8 \pi ^2 \left(-8 m_K^2+m_{\pi }^2+2 \left(m_{\eta }^2+m_{\eta
        '}^2\right)\right)+3 f_{\pi }^2 \left(5 g^4 N_c-136 \pi ^2 \lambda _1\right)\right)+\nn\\
       &&f_{\pi}^3 \left(16 \pi ^2 \left(9 m_K^2+16 m_{\pi }^2-8 \left(m_{\eta
       }^2+m_{\eta'}^2\right)\right)\right.+\nn\\
      && \left.\left.\left.f_{\pi }^2 \left(-9 g^4 N_c+384 \pi ^2 \lambda
      _1\right)\right)\right)\right)/\left(96 \pi ^2 f_K \left(8 f_K^3-16 f_K^2 f_{\pi }+11 f_K
       f_{\pi }^2-3 f_{\pi }^3\right)\right) ;\nn\\
m_{s,08}^2&=&\left(8 f_K^4 \left(g^4 N_c-16 \pi ^2 \lambda _1\right)-16 f_K^3 f_{\pi } \left(g^4 N_c-12 
\pi ^2 \lambda _1\right)+8 \pi ^2 f_{\pi }^2 \left(2 m_{\pi }^2-m_{\eta }^2-m_{\eta '}^2+3 f_{\pi }^2 
\lambda _1\right)+\right.\nn\\
         &&f_K^2 \left(32 \pi ^2 \left(-4 m_K^2+2 m_{\pi }^2+m_{\eta }^2+m_{\eta '}^2\right)+f_{\pi }^2 
\left(11 g^4 N_c-48 \pi ^2 \lambda _1\right)\right)-\nn\\
         &&\left.f_K f_{\pi } \left(8 \pi ^2 \left(-7 m_K^2+m_{\pi }^2+3 \left(m_{\eta }^2+m_{\eta '}^2\right)\right)+
     f_{\pi }^2 \left(3 g^4 N_c+40 \pi ^2 \lambda _1\right)\right)\right)/\left(6 \sqrt{2} \pi ^2 \left(8 f_K^2-
     8 f_K f_{\pi }+3 f_{\pi }^2\right)\right) ;\nn\\
\tan\l2\theta_s\r&=&\frac{2m^2_{s,08}}{m^2_{s,00}-m^2_{s,88}} ;
\label{parametersall}
\eeqa
Hence only $\lambda_2$ depends on $\Lambda$. It is straightforward but tedious to check that this dependence cancels neatly
with that of $\Omega_{\bar{q}q}^{\text{v}}$ in (\ref{eq.omega}) to yield a $\Lambda$ independent $\Omega$.

\end{document}